\documentclass[sigconf]{acmart}
\definecolor{highlight}{RGB}{0, 0, 150}
\usepackage{algorithm}
\usepackage{algpseudocode}
\usepackage{makecell}
\usepackage{multirow}
\usepackage{caption}
\usepackage{subcaption}
\usepackage{xcolor}
\usepackage{soul}
\usepackage{balance}
\usepackage{titlesec}
\titlespacing*{\section}{0pt}{4pt plus 1pt}{4pt minus 1pt}
\titlespacing*{\subsection}{0pt}{4pt plus 1pt}{4pt minus 1pt}

\AtBeginDocument{%
  \providecommand\BibTeX{{%
    \normalfont B\kern-0.5em{\scshape i\kern-0.25em b}\kern-0.8em\TeX}}}

\copyrightyear{2023} 
\acmYear{2023} 
\setcopyright{rightsretained} 
\acmConference[WWW '23]{Proceedings of the ACM Web Conference 2023}{May 1--5, 2023}{Austin, TX, USA}
\acmBooktitle{Proceedings of the ACM Web Conference 2023 (WWW '23), May 1--5, 2023, Austin, TX, USA}
\acmDOI{10.1145/3543507.3583420}
\acmISBN{978-1-4503-9416-1/23/04}

\begin{document}
\title{Zero-shot Clarifying Question Generation for Conversational Search}

\author{Zhenduo Wang}
\email{zhenduow@cs.utah.edu}
\orcid{1234-5678-9012}
\affiliation{%
  \institution{University of Utah}
}

\author{Yuancheng Tu}
\email{yuantu@microsoft.com}
\orcid{1234-5678-9012}
\affiliation{%
  \institution{Microsoft}
}

\author{Corby Rosset}
\email{corbyrosset@microsoft.com}
\orcid{1234-5678-9012}
\affiliation{%
  \institution{Microsoft}
}

\author{Nick Craswell}
\email{nickcr@microsoft.com}
\orcid{1234-5678-9012}
\affiliation{%
  \institution{Microsoft}
}

\author{Ming Wu}
\email{mingwu@microsoft.com	}
\orcid{1234-5678-9012}
\affiliation{%
  \institution{Microsoft}
}

\author{Qingyao Ai}
\email{aiqy@tsinghua.edu.cn}
\affiliation{%
  \institution{Tsinghua University}
  }


\begin{abstract}

A long-standing challenge for search and conversational assistants is query intention detection in ambiguous queries. 
Asking clarifying questions in conversational search has been widely studied and considered an effective solution to resolve query ambiguity. 
Existing work have explored various approaches for clarifying question ranking and generation. 
However, due to the lack of real conversational search data, they have to use artificial datasets for training, which limits their generalizability to real-world search scenarios.
As a result, the industry has shown reluctance to implement them in reality, further suspending the availability of real conversational search interaction data.
The above dilemma can be formulated as a cold start problem of clarifying question generation and conversational search in general.
Furthermore, even if we do have large-scale conversational logs, it is not realistic to gather training data that can comprehensively cover all possible queries and topics in open-domain search scenarios.
The risk of fitting bias when training a clarifying question retrieval/generation model on incomprehensive dataset is thus another important challenge.

In this work, we innovatively explore generating clarifying questions in a zero-shot setting to overcome the cold start problem and we propose a constrained clarifying question generation system which uses both question templates and query facets to guide the effective and precise question generation. The experiment results show that our method outperforms existing state-of-the-art zero-shot baselines by a large margin. Human annotations to our model outputs also indicate our method generates 25.2\% more natural questions, 18.1\% more useful questions, 6.1\% less unnatural and 4\% less useless questions.

\end{abstract}


\settopmatter{printfolios=true}
\maketitle

\section{Introduction}
One common cause of search failure is ambiguity in queries, which refers to the queries with multiple relevant information needs or unclear intent. 
Ambiguous queries are often the result of users not knowing how to formulate their needs. 
For example, a user looking for "the Discovery Channel's dinosaur site with pictures and games of dinosaurs" and a user looking for "different kinds of dinosaurs" can both search with the query "dinosaur".
Ambiguous queries can also indicate the user is conducting an exploratory search, such as learning or investigating searches \cite{marchionini2006exploratory}.
A popular system feature for query ambiguity is search result page diversification \cite{santos2015search, keyvan2022approach}. However, it can hardly be applied to searches on devices with small screens or devices with only speech functions by design.

In fact, both scenarios of ambiguous query incentivize the search system to have multi-turn user-system interaction capabilities, i.e., Conversational Search, which has recently become a growing research frontier in the Information Retrieval (IR) community. 
Conversational Search addresses query ambiguity by arguably its most characterizing feature, mix-initiative interactions. 
It means that not only the user but also the system can proactively lead the conversation by asking clarifying questions about the user's search intent. 
These clarifying questions chiefly determine the quality of conversational search.
Therefore, existing works have extensively explored various approaches to selecting or generating high-quality clarifying questions.

However, there are two challenges that limits the application of conversational search systems in real-world. First, as a new retrieval paradigm, there isn't any mature online service for open-domain conversational search. The cost of collecting large-scale conversational search logs is still prohibitive, and the building and evaluation of reliable conversational systems is thus difficult, which further increase the difficulty of collecting conversational logs in practice. We refer to this dilemma as the cold-start problem for clarifying question generation. Second, traditional clarifying question generation methods \cite{zamani2020generating, sekulic2021towards} often rely on supervised learning with labeled or artificial conversational logs. It is unrealistic to require such logs to cover all the topics of possible queries, and models trained with incomprehensive datasets could suffer from catastrophic forgetting \cite{mccloskey1989catastrophic} and inevitably be biased on unseen queries. We refer to this problem as the data bias in conversational search log collection.




Unlike previous studies \cite{rao2019answer, zamani2020generating, dhole2020resolving, sekulic2021towards, wang2021template}, we explore a new task of clarifying question generation in zero-shot scenarios without the use of conversational search logs. The main idea is to learn a clarifying question generation model directly from large-scale text data and search engine traffic without collecting or labeling conversational data from a conversational search system. In this way, we can avoid both the cold start and data bias problem from the very beginning. While there have been many studies on zero-shot text generation, as shown in this paper, applying these methods to clarifying question generation directly often produce unsatisfactory results because of two reasons: First, the conventional sequence-to-sequence language generation models cannot efficiently learn the needed correlations between the initial queries submitted by users and the clarifying questions generated by systems. Their generations tend to talk about general topics that are not relevant to the specific search need; Second, existing zero-shot language model generations are usually narratives instead of questions. How to guide the zero-shot model to generate text in question forms that are proper for each search query is still unknown.


To solve the above problem, in this paper, we propose to constrain the clarifying question decoding with search facets. 
Facets refer to possible subtopics of a query (e.g., "pictures", "map", "populations" are possible facets for the query "I am looking for information about South Africa") that can be effectively extracted from search result pages (SERP) \cite{zhao2022generating}, knowledge graphs \cite{xu-etal-2019-asking}, or other sources \cite{rao2019answer, white-etal-2021-open} in unsupervised manners.
Constrained language decoding refines the naive beam-search decoding with the ability to rank facet-containing generations higher, resulting in more questions about the facet.
We also initialize the decoding with questioning prompts instead of generating the entire question sentence.
Multiple question templates are used in this process and eventually ranked for the best generation.

To demonstrate the effectiveness of our zero-shot system, we compare with multiple existing non-facet and facet-driven baselines, including several state-of-the-art supervised learning methods.
They will be finetuned on a training set, which is not accessible by our zero-shot system.
Nonetheless, we show that our system significantly improves these baseline methods by a large margin, which implies our system is the best solution for zero-shot clarifying question generation. 

During the evaluation phase, we compute automatic metrics \cite{papineni-etal-2002-bleu, lin-2004-rouge, banerjee-lavie-2005-meteor, lin-etal-2020-commongen} and employ humans to provide quality labels for the generations of the compared systems.
Our human annotators evaluate the generations from both their \textit{naturalness} \cite{sekulic2021towards} from language perspective and \textit{usefulness} \cite{sekulic2021towards, rosset2020leading} from utility perspective. 
The automatic metric scores are our primary evaluation, from which we conclude that our system is the best. 
Human annotation results suggest our method generates 25.2\% more natural questions, 18.1\% more useful questions, 6.1\% less unnatural and 4\% less useless questions, which aligns with automatic evaluations and reinforces our conclusions with confidence.

We consider the following key contributions of our work:

\vspace{-5pt}
\begin{itemize}
    \item We are the first to propose a \textbf{zero-shot} clarifying question generation system, which attempts to address the cold-start challenge of asking clarifying questions in conversational search. 
    The zero-shot setting also maximizes the generalizability of our system to serve different search scenarios.
    \item We are the first to cast clarifying question generation as a constrained language generation task and show the advantage of this configuration.
    We show that a simple constrained decoding algorithm, even under zero-shot setting, can guide clarifying question generation better than finetuning the model with limited training data.
    Our work is a compelling demonstration of how large deep models benefit from properly integrating human knowledge.
    \item We propose an auxiliary evaluation strategy for clarifying question generations, which removes the information-scarce question templates from both generations and references.  
    Results computed this way expose the limitations of the existing default evaluation strategy and provide insights into the actual quality of generated questions.
\end{itemize}

\section{Related Works}
\subsubsection*{\textbf{Conversational Search}}
Conversational Search refers to the process of information-seeking involving natural language conversations with the search system \cite{zamani2022conversational}. 
It has been identified as one of the most important research area of IR \cite{anand2020conversational, culpepper2018research}.
Recently, a plethora of seminars and tutorials has been given about Conversational Search from different standpoints such as \cite{anand2020conversational,hauff2021conversational, fu2020tutorial, gao2018neural, zamani2022conversational, gao2022neural}. 
\citeauthor{radlinski2017theoretical} \cite{radlinski2017theoretical} proposed a theoretical framework for Conversational Search and highlighted mix-initiative as one of its most desired perspective.
Later, \citeauthor{zamani2020macaw} \cite{zamani2020macaw} designed an abstract pipeline for a complete conversational search system. 
Conversational Search is also closely related to many other research areas such as Task-oriented Dialog System, Conversational Question Answering, Conversational Recommendation, and Chatbot. 
\citeauthor{anand2020conversational} \cite{anand2020conversational} provided one possible view to connect Conversational Search, Dialog System, and Chatbot.  
Recently, \citeauthor{zamani2022conversational} \cite{zamani2022conversational} defined Conversational Search, Recommendation and Question Answering as subdomains of Conversational Information Seeking.

\subsubsection*{\textbf{Resolving Query Ambiguity}}
The query ambiguity problem is an important motivation to promote conversational search over conventional single-turn search. 
Ambiguous query generally refers to the queries for which the search system cannot confidently identify the user's information need and return search results \cite{keyvan2022approach}. 
Queries can be ambiguous for various reasons, such as containing multiple distinct interpretations or under-specified subtopics \cite{Clarke2009OverviewOT}, anaphoric ambiguity, and syntactic ambiguities \cite{schlangen2004causes}. 
Approaches to clarifying query ambiguity can be roughly divided into three categories: (1) Query Reformulation such as \cite{liu2010analysis, dang2010query, elgohary2019can, yu2020few} iteratively refines the query; (2) Query Suggestion such as \cite{sordoni2015hierarchical, rosset2020leading, yang2017neural} offers related queries to the user; (3) Asking Clarifying Questions such as \cite{zamani2020generating, zhang2018towards, bi2021asking, rao2018learning, rao2019answer} proactively engages users to provide additional context. 
While the three approaches share many structural and functional similarities, they cannot be replaced by each another. 
Because none of them is the best in all scenarios, for example, asking clarifying questions could be exclusively helpful to clarify ambiguous queries without context. In contrast, query reformulation is more efficient in context-rich situations. 
Query suggestion is good for leading search topics, discovering user needs, etc.

\subsubsection*{\textbf{Asking Clarifying Questions}}
Among the approaches to resolving query ambiguity, asking clarifying questions (CQ) is the most studied \cite{keyvan2022approach}, and is considered as more convenient because of its proactivity \cite{zhao2022generating, radlinski2017theoretical, vtyurina2017exploring}.
Existing studies about asking CQ can be divided into two main categories: (1) ranking/selecting CQ such as \cite{aliannejadi2020convai3, bi2021asking, rao2018learning}, and (2) generating CQ such as \cite{rao2019answer, zamani2020generating, dhole2020resolving, sekulic2021towards, wang2021template}. \citeauthor{rao2019answer} \cite{rao2019answer} applied generative adversarial learning in training sequence-to-sequence question generation model. \citeauthor{zamani2020generating} \cite{zamani2020generating} proposed a rule-based template completion model and two neural question generation models to generate CQ given the query and its aspect. 
Later in \cite{dhole2020resolving, wang2021template}, the authors also demonstrated templates could guide CQ generation. Their solutions effectively convert the CQ generation problem to a selection task. 
Similar to using query aspect, \citeauthor{sekulic2021towards} \cite{sekulic2021towards} also proposed a query facet-driven approach. 
Recently, \citeauthor{zhao2022generating} \cite{zhao2022generating} showed such query facets could be extracted from web search results and guide question generation.

\subsubsection*{\textbf{Constrained Natural Language Generation}}
Our work applies constrained natural language generation to generating clarifying questions for conversational search. 
The task of constrained natural language generation was proposed in \cite{anderson-etal-2017-guided}, where the problem was modeled as beam search over $2^{C}$ states representing all combinatorial satisfaction states of $C$ constraints. 
This exponential complexity limited its applications. 
\citeauthor{hokamp-liu-2017-lexically} \cite{hokamp-liu-2017-lexically} propose a grid beam search method that groups beams by the number of constraints already satisfied. 
\citeauthor{miao2019cgmh} \cite{miao2019cgmh} propose to edit generations using constraints with Metropolis Hastings sampling. 
\citeauthor{welleck2019non} \cite{welleck2019non} develop a non-monotonic tree-based generation system which can generate texts given constraints at arbitrary positions. 
\citeauthor{zhang-etal-2020-language-generation} \cite{zhang-etal-2020-language-generation} suggest a tree-enhanced Monte-Carlo approach for text generation via Combinatorial Constraint Satisfaction. 
More recently, \citeauthor{lu2020neurologic} \cite{lu2020neurologic, lu2021neurologic} propose NeuroLogic Decoding and A$^\star$ search. 
Their decoding algorithms incorporate constraints as Conjunctive Normal Forms (CNF) and estimate the viability of each beam to satisfy constraints by sampling their future generations.

\section{Zero-shot Facet-constrained Question Generation}

This section gives detailed descriptions of our zero-shot clarifying question generation system, which addresses two challenges in naive models for zero-shot clarifying question generation.
Our system is zero-shot, meaning that we do not train our system on any training data for clarifying question generation. 
The generation is also facet-constrained, which implies that we use the search facet in our question generation. 
A facet is one possible search direction for the ambiguous query; for example, \textit{pictures, map, location, populations} are possible facets for the query "I am looking for information about South Africa".
Facet has been considered useful \cite{zamani2020generating,sekulic2021towards, zhao2022generating} for clarifying question generation since it provides a relevant direction for inquiring about the user intent.
Clarifying question generation can be challenging without facets because the generations are often too general and clueless.
In \cite{zhao2022generating}, \citeauthor{zhao2022generating} proposes a facet extraction approach, which shows that these useful keywords can be easily extracted from web search results. 
Previous works also suggest that facets can also be extracted from various sources, including product reviews \cite{rao2019answer}, images \cite{white-etal-2021-open}, or knowledge graphs \cite{xu-etal-2019-asking}.

The backbone of our system is a checkpoint of the public Generative Pretrained Transformers (GPT-2) \cite{radford2019language} pretrained on a seperate large scale text corpus. 
Originally, the inference objective of GPT-2 is to predict the next token given all previous texts.
\begin{equation}
    L = \sum_{t=1}^T \log P(x_t|x_{1:(t-1)}, \theta)
\end{equation}

where $x_{1:(t-1)}$ is the generated sequence by step $(t-1)$, $\theta$ is GPT-2 parameters. 

One naive method to adapt GPT-2 for clarifying question generation is to append the query $q$ and facet $f$ together as initial texts and let GPT-2 generate a continuation $cq$ as the clarifying question. 
However, this method faces two challenges. 
The first challenge is it does not necessarily cover facets in the generation.
Previous work \cite{sekulic2021towards} proposes a finetuning approach, which trains on a collection of $f \text{ [SEP] } q \text{ [BOS] } cq \text{ [EOS] } $ paragraphs.
However, as reported in their work, this structure does not outperform simply using the query alone as input. 
We analyze the generations of their model and find that the coverage of facet words in these generations is only about 20\%. 
This number implies that simply appending facet words to the input of GPT-2 is highly inefficient in informing the decoder. 
    
The other challenge is that the generated sentences are not necessarily in the tone of clarifying questions. 
This is because clarifying questions makes up only a small portion of natural language usage. GPT-2 is pre-trained on web texts, most of which are narrative. 
Even for the questions, they are not necessarily for the purpose of clarifying. 
As a result, pre-trained GPT-2 often generates relevant factoids following the query and facet.

To explain our proposed system easier, we divide our system into two parts: (1) facet-constrained question generation and (2) multiform question prompting and ranking.
The two parts respectively address the above two challenges of zero-shot GPT-2.

\begin{table*}[t]
\caption{The most common 4-grams in clarifying question answers \cite{krasakis2020analysing} and their corresponding questions with example generations from GPT-2 for the query "I am looking for information about South Africa." Duplicated Examples are omitted.}
\vspace{-10pt}
\begin{tabular}{l|l|l|l|l|l}

\toprule
     1st word             & 2nd word            & 3rd word              & 4th word    & Corresponding question  & Example\\ \hline
  \multirow{2}{*}{Yes} & \multirow{2}{*}{I}  & would                 & like        & would you like to  &  would you like to [take pictures of]    \\ \cline{3-6} 
                          &                     & want                  & to          & do you want to  &   do you want to [see pictures of them]           \\ \cline{1-6} 
     \multirow{11}{*}{No} & \multirow{10}{*}{I} & \multirow{2}{*}{am}   & interested  & are you interested in  &  are you interested in [taking pictures of them]  \\ \cline{4-6} 
                          &                     &                       & looking     & are you looking for    &  are you looking for [pictures of South Africa]  \\ \cline{3-6} 
                          &                     & \multirow{2}{*}{just} & want        & do you want to      &  ...   \\ \cline{4-6} 
                          &                     &                       & need        & do you need to      &  do you need to [send pictures to us]             \\ \cline{3-6} 
                          &                     & \multirow{2}{*}{need} & to          & do you need to      &  ...  \\ \cline{4-6} 
                          &                     &                       & information & do you need information  & do you need information [or pictures of South Africa] \\ \cline{3-6} 
                          &                     & \multirow{3}{*}{want} & information & do you want information & do you want information [and pictures of South Africa] \\ \cline{4-6} 
                          &                     &                       & the         & do you want to      &  ...   \\ \cline{4-6} 
                          &                     &                       & to          & do you want to      &  ...   \\ \cline{3-6} 
                          &                     & would                 & like        & would you like to   &  ...   \\ \cline{2-6} 
                          & Im                  & looking               & for         & are you looking for   & ...  \\ \cline{1-6} 
     \multirow{4}{*}{I}   & am                  & looking               & for         & are you looking for   & ...  \\ \cline{2-6} 
                          & dont                & know                  & .           & do you want to know   & do you want to know [about the pictures and videos] \\ \cline{2-6} 
                          & want                & to                    & know        & do you want to know   & ... \\ \cline{2-6} 
                          & would               & like                  & to          & would you like to     & ... \\ 
                        
\bottomrule
\end{tabular}
\vspace{-10pt}

\label{table:common4gram}
\end{table*}


\subsection{Facet-constrained Question Generation}
\label{sec:decoding}
In the first abovementioned challenge, we find that existing works struggle to generate facet-related clarifying questions.
Unlike these works, we believe that simply appending the facet to the input is inefficient. Instead, our model utilizes the facet words as decoding constraints. 
Specifically, we see the task of generating facet-related questions as a facet-constrained language generation problem, i.e., to use the facet words as constraints during generation decoding. 
To encourage the decoder to choose generations containing more facet words, we employ an algorithm called NeuroLogic Decoding \cite{lu2020neurologic}.

NeuroLogic Decoding is based on beam-search decoding. 
In each decoding step $t$, assuming the already generated candidates in the beam are $C = \{c_{1:k}\}$, where $k$ is the beam size, $c_i = x_{1:(t-1)}^i$ is the $i$th candidate, and $x_{1:(t-1)}^i$ are tokens generated from decoding step $1$ to $(t-1)$, NeuroLogic Decoding works in the following steps:

\begin{enumerate}
    \item Generate the next token distributions $x_{t}^i\sim P(x_{1:(t-1)}^i)$ for each candidate in the beam with GPT-2. Assume that the vocabulary size is $|V|$, then this will create $k\times |V|$ new candidates.
    
    \item \textbf{Pruning Step:} Among these candidates, discard all but candidates that are in both in the top-$\alpha$ tokens in terms of $p(x_{1:t})$ and the top-$\beta$ in terms of number of facet words contained by $x_{1:t}$. 
    \label{step:top_ab_filter}
    
    \item \textbf{Grouping Step:} Group the remaining candidates by the facet words they contain. 
    This will result in $2^{|f|}$ groups. However, notice that some group could be empty.
    \label{step:grouping}

    \item Keep the best candidate from each of the groups. 
    Again, there may be less than $2^{|f|}$ candidates left now. 
    Just keep at most the best $k$ candidates with the highest $p(x_{1:t})$ in the beam and move onto decoding the $(t+1)$th token.
    
\end{enumerate}

We recommend a more vivid demonstration in the original paper \cite{lu2020neurologic}. We now explain why NeuroLogic Decoding could better constrain the decoder to generate facet-related questions by highlighting some key steps.
First, the top-$\beta$ filtering in step~(\ref{step:top_ab_filter}) is the main reason for promoting facet words in generations. 
Because of this filtering, NeuroLogic Decoding tends to discard generations with fewer facet words regardless of their generation probability. 
Therefore, facet-related generations with low probability will more likely stand out against greedy high-probability generations without using facet words. 
Then, the grouping in step~(\ref{step:grouping}) is the key for NeuroLogic Decoding to explore as many branches as possible. 
Because this grouping method keeps the most cases ($2^{|f|}$) of facet word inclusions, allowing the decoder to cover the most possibilities of ordering constraints in generation. 

As mentioned in \cite{lu2020neurologic}, the asymptotic runtime of NeuroLogic Decoding is $O(NK)$, where $N$ is the text sequence length and $k$ is beam search size. 
This is the same as normal beam search and faster than most previous constrained language generation algorithms, making it fairly applicable in real cases.

\subsection{Multiform Question Prompting and Ranking}
\label{sec:ranking}

Another challenge is guiding zero-shot GPT-2 to generate clarifying questions instead of narrative or other types of questions. 
We use clarifying question templates as the starting text of the generation and let the decoder generate the rest of question body. 
For example, if the query is "I am looking for information about South Africa." 
Then we give the decoder "I am looking for information about South Africa. [SEP] would you like to know" as input and let it generate the rest.
From our observation, GPT-2 is much better at finishing a question like this than asking a new question by itself.

In our system, we use multiple prompts because we want to both cover more ways of clarification with different prompts and avoid making users bored with monotonic questions. 
A previous study about the effect of clarifying question \cite{krasakis2020analysing} shows the most common 4-grams for answering clarifying questions, as shown in Table~\ref{table:common4gram}. 
Inspired by their work, we reverse these most common answers to their original question forms (eight in total) and use them as our prompt candidates. 
For each query, we will append these eight prompts to the query and form eight inputs. 
Eventually, we will generate eight clarifying question candidates. 
For example, our generated questions for the query "I am looking for information about South Africa." with facet "map" is shown in Table~\ref{table:common4gram}.

In real applications, our system should return one question in the form of the concatenation of the prompt and the GPT-2 output. 
To find the best question, we explore various ranking methods to rank our prompted generations:

\subsubsection*{\textbf{Perplexity}} is a commonly used method that ranks clarifying questions by the perplexity of query-question concatenation computed by pre-trained and GPT-2.

\subsubsection*{\textbf{AutoScore}} is another commonly used method that ranks clarifying questions by weighted sum of automatic natural language generation scores including BLEU \cite{papineni-etal-2002-bleu}, ROUGE \cite{lin-2004-rouge}, and METEOR \cite{banerjee-lavie-2005-meteor}. 
These scores are computed using generated questions as hypotheses and queries as references.

\subsubsection*{\textbf{Cross-encoder}} \cite{humeau2019poly} is a typical and commonly used dense retrieval structure. 
It ranks question candidate by its relevance which is computed by a transformer encoder followed by a linear scoring layer. 
The cross-encoder is pre-trained on millions of Reddit dialogues \cite{mazare2018training}. 
Directly using the pre-trained checkpoint is potentially suboptimal because the prompted generation ranking objective differs from the pretraining task. 

\subsubsection*{\textbf{NTES}} \cite{ou2020clarifying} is a clarifying question ranking model that wins the ConvAI3 challenge \cite{aliannejadi2020convai3} on ClariQ dataset. 
The clarifying question ranking subtask in ConvAI3 requires a system to rank clarifying question candidates given query and facet. 
We consider this task highly similar to our prompted generation ranking task. 
The NTES model finetunes pretrained ELECTRA \cite{clark2020electra} as its ranker on ClariQ dataset. 
We use their finetuned checkpoint and aim to leverage the clarifying question ranking knowledge in this model. 

\subsubsection*{\textbf{Weighted Sequential Dependency Model (WSDM)}} \cite{bendersky2010learning} is a document ranking method based on \emph{query-candidate} overlap in terms of unigram, and ordered/unordered bigram within a context window. 
Our system treats the center words in the original query together with facet words as \emph{query} and prompted generation as \emph{candidates}.
The motivation of WSDM is to rank those generations with facet words co-occurring together higher. 
For example, given the query "I'm looking for information about South Africa" and facet "map", the WSDM model will rank all the prompted questions using "South Africa map" as the query. 
Questions like "do you need information about the map of South Africa" will be ranked higher than "do you want to buy a map that is made in South Africa" because "South Africa" and "map" are closer in the first question and more likely to be more useful clarifications.

We have empirically compared different versions of our model using the ranking methods above in the experiments and find that WSDM achieves the best empirical performance overall. Thus, if not mentioned, we use WSDM as our primary ranker. 

\section{Experiments}

\subsection{Research Questions and Experiment Design}
We design experiments to answer the following research questions:

\subsubsection*{\textbf{RQ1.}} How well can we do in zero-shot clarifying question generation with existing baselines?

In this research question, we show the performance of some clarifying question generation baselines in the zero-shot setting: 
\begin{enumerate}
    \item \textbf{Q-GPT-0} simply uses GPT-2 to generate the clarifying question given the query. This and the next baseline are the zero-shot version of approaches in \cite{sekulic2021towards}.
    
    \item \textbf{QF-GPT-0} appends the facet to the front of the query and generates the clarifying question. 
    
    \item \textbf{Prompt-based GPT-0} is a prompt-based GPT-2 approach which includes a special instructional prompt as input:
    
    $q$ "Ask a question that contains words in the list [$f$]."
    
    \item \textbf{Template-0} is a template-guided approach using GPT-2. As mentioned earlier, a common problem for zero-shot GPT-2 is that it mostly generates narratives instead of questions. The Template baseline add the eight question templates during decoding and generate the rest of the question, which is similar to approaches in \cite{dhole2020resolving, wang2021template}.

\end{enumerate}

\subsubsection*{\textbf{RQ2.}} How effective is facet information for clarifying question generation if utilized efficiently? 

To the best of our knowledge, no previous work has explored clarifying question generation using the ambiguous query as the only source of information. 
Previous works such as \cite{zamani2020generating, sekulic2021towards, xu-etal-2019-asking} propose various facet-specific clarifying question generation methods using facet or aspect of the query. 
Despite this, most of them did not or failed to experimentally demonstrate the importance of additional information for facet-specific clarifying question generation. 
Particularly in \cite{sekulic2021towards}, it is shown that adding facet does not significantly improve the quality of generated questions. 
These works leave the effectiveness of facet in doubt.
We argue that the way facet information is utilized in these works is inefficient.

To answer this research question, we compare our proposed zero-shot facet-constrained approach with a facet-free variation that uses subject words from the query as constraints. 
For example, the subject words of the query "I am looking for information about South Africa." is "South Africa". 
Using a part-of-speech tagger, we extract the nouns or proper nouns as subject from the query.
    
\subsubsection*{\textbf{RQ3.}} How does our zero-shot facet-constrained approach compare to existing facet-driven baselines?

    
To answer this research question, we include some existing methods and a few other reasonable solutions not mentioned by previous works as our baseline models.
Some of them are zero-shot, while others are not. 
However, we still compare their performances jointly to demonstrate our zero-shot approach's power. We divided the dataset into training and evaluating sets. All the finetuning methods can access the training set to finetune pre-trained GPT-2 checkpoint, while our zero-shot system cannot access them. Then, we evaluate all the methods on the evaluation set. 

We compare our model against the following baseline models:

\begin{enumerate}
    \item \textbf{Template-facet} is a clarifying question rewriting baseline which appends the facet word right after the question template.
    For a fair comparison, we also apply multiform question templates and ranking. 
    For example, given query $q$ and facet $f$, we first generate eight questions by appending facet to each of the eight templates in Table~\ref{table:common4gram}. 
    Then we rank these questions by their language perplexity.
    This baseline is not ideal. Admittedly, it can generate good questions such as: 
    \begin{enumerate}
        \item[] $q$: "I am looking for information about South Africa."
        \item[] $f$: "population"
        \item[] $cq$: "Are you interested in [population]" 
    \end{enumerate}
    
    However, sometimes the facet itself is not meaningful:
    
    \begin{enumerate}
        \item[] $q$: "I am interested in poker tournaments."
        \item[] $f$: "online"
        \item[] $cq$: "Are you interested in [online]"
    \end{enumerate}
    
    \item \textbf{QF-GPT} \cite{sekulic2021towards} is a GPT-2 finetuning version of \textbf{QF-GPT-0}. 
    It initializes with pretrained GPT-2 and finetunes on a set of (facet $f$, query $q$, clarifying question $cq$) tuples in the form as $f \text{ [SEP] } q \text{ [BOS] } cq \text{ [EOS] } $ paragraphs, where [SEP] is the separator token, [BOS] the beginning-of-sentence token, and [EOS] the end-of-sentence token. 
    
    \item \textbf{Prompt-based finetuned GPT} is a finetuning version of \textbf{Prompt-based GPT-0}
    The motivation is that simple facet-as-input finetuning is highly inefficient in informing the decoder to generate facet-related questions by observing a facet coverage rate of only 20\%.
    Inspired by recent advances in prompt studies, especially for natural language generation such as \cite{liu2021pre, schick2021few, li2021prefix}, we add a sentence "Ask a question that contains words in the list [$f$]" between $q$ and $cq$, aiming to instruct GPT-2 the inclusion of facet words in the clarifying question. Hence, we finetune GPT-2 with the structure: 
    
    $q$ "Ask a question that contains words in the list [$f$]." $cq$

\end{enumerate}

    

\subsection{Dataset}
We use ClariQ-FKw \cite{sekulic2021towards} dataset for our main experiments. 
ClariQ dataset is originally from ConvAI3 challenge \cite{aliannejadi2020convai3}. 
This dataset has rows of $(q, f, cq)$ tuples, where $q$ is an open-domain search query, $f$ is a search facet, and $cq$ is a human-generated clarifying question regarding the facet. 
The facet in ClariQ is in the form of a faceted search query. 
ClariQ-FKw extracts the keyword of the faceted query as its facet column and samples a dataset with 1756 training examples and 425 evaluation examples. 
We report the performances of all of our proposed and baseline systems on its evaluation set. 
Because we aim to solve the problem in a zero-shot setting, our proposed system does not access the training set. 
The other supervised learning systems can access the training set for finetuning.

\subsection{Evaluation}
\label{sec:evaluation}

We use automatic metrics for natural language generation and human annotators to evaluate system performances to label the generated questions. 
Following previous works \cite{sekulic2021towards, aliannejadi2020convai3, aliannejadi2019asking}, we use the human generated question from the dataset as gold reference.
It is worth to clarify that this type of evaluation and its metrics are only meant to measure the the ability of a system to ask clarifying questions about the facet specifically, instead of generally relevant questions to query.
However, defining other types of evaluation will be challenging given what we have in the dataset.

\subsubsection{Automatic Metrics} 

The automatic metrics we use are BLEU  \cite{papineni-etal-2002-bleu}, ROUGE \cite{lin-2004-rouge}, METEOR \cite{banerjee-lavie-2005-meteor}, and Coverage \cite{lin-etal-2020-commongen}. 
BLEU and ROUGE are based on word match, while METEOR uses more general word forms to compute alignments between reference and generation. 
Coverage is computed as the average frequency of facet words in generations. 
Because these automatic metrics are mostly based on word overlap, we propose two different ways of computing them. 
We argue that not all the words in the generation are equally important. 
Such as the words in the question templates are less important than those in the actual question body. 
For example:

\begin{enumerate}
    \item[] $q$: "I am looking for information about South Africa."
    \item[] $f$: "picture"
    \item[] \textit{ref}: "would you like to [see some pictures of South Africa]"
    \item[] \textit{cq1}: "\ul{would you like to} [take \ul{\textbf{pictures of}}]"
    \item[] \textit{cq2}: "are \ul{you} looking for [\ul{\textbf{pictures of South Africa}}]"
\end{enumerate}

In this example, \textit{ref} is the gold reference clarifying question, \textit{cq1} and \textit{cq2} are two candidate generations. 
Underline means word overlap between candidates and references. 
\textit{cq1} has more word overlaps than \textit{cq2}, including a 4-gram overlap "would you like to" with the reference. 
However, we humans can quickly tell that \textit{cq1} is a worse clarifying question than \textit{cq2}. 
The reason for this discrepancy is that the question template does not contain real information. 
Therefore it can easily bias generations with the same template but a bad question body. 
We are concerned that the conventional evaluation approaches computed on full questions could have limitations in our task. 
Plus the fact that the question templates are not actually `generated' by the models. (They are given.)
Hence, we propose another way to compute these metrics without the question template and on the actual question body. 
In the example, we bold the word overlap between candidates and reference within the question body. \textit{cq2} has more overlap than \textit{cq1} in this way, which corresponds to the human judgment of good and bad clarifying questions.

After reading Section~\ref{sec:decoding} and \ref{sec:evaluation}, a reasonable concern would be: wouldn't the facet-constrained generation naturally improve the automatic metrics?
Because using facets as constraints will make these true-positive words more likely to be included.
We are aware of this concern, and we design multiple experiments and evaluations to ensure the performances of our system are meaningful. 
First, in RQ3, we compare our system with the Template-facet rewriting baseline, which benefits even more than our system because of the guaranteed facet inclusion.
We will show that our proposed system can achieve even higher scores than Template-facet in Section~\ref{sec:results}. 
Second, we include human evaluations. Human annotators are free of this bias because they will evaluate generated questions by the quality of entire sentences, not word overlaps.
Last, we want to highlight that facet is used in one way or another by all facet-driven models in RQ3. 
Using them as constraints or inputs is a modeling choice that does not break fair comparison principles.

\subsubsection{Human Evaluation Metrics} Like the example above, automatic metrics are reported \cite{belz2006comparing} for not necessarily corresponding to true generation quality. 
Therefore, following previous works \cite{zamani2020generating, sekulic2021towards, rosset2020leading}, we employ human annotators to evaluate the generated clarifying question qualities on 425 test examples.
The annotators are provided randomly shuffled generations from all the models in RQ3 and asked to label them without knowing their sources.
We provide the annotators with a detailed guideline to annotate the generated question into two labels: usefulness and naturalness. 
For each label, the annotators must decide whether the question is good, fair, or bad. 
The guideline can be found in appendix~\ref{appendsec:guide}.

\subsubsection*{\textbf{Naturalness}} is defined as the general fluency and understandability of the generated question. 
The naturalness of a question is independent of its coherence to the topic of the query. 
By our definition, this label mainly evaluates the overall language modeling capacities of the model.
Generally speaking, a zero-shot GPT-2-based decoder would keep the same language capacity as the original GPT-2 because it uses the same model. 
However, finetuning could downgrade the capacity due to the bias of the limited-sized finetuning set. 
    
\subsubsection*{\textbf{Usefulness}} is defined as whether the question is relevant to both the query and the facet and makes the query easier to answer.
Typical bad usefulness questions can fall into one of the categories: duplicate, prequel, miss-intent, too general, or too specific \cite{rosset2020leading}.
These questions are relevant to the query but not useful for clarification.
For example, for the query "Tell me about computer programming." and facet "courses", "are you interested in computer programming." is a duplicate question with the original query, and "are you looking for computer programming courses for children." is too-specific.

\begin{table*}[t]
\caption{Model performances evaluated on full question and reference. ZSFC models are our zero-shot facet-constrained method with different question rankers. Bolded numbers indicate the best-performing model of the column.\protect\footnotemark}
\vspace{-7pt}
\begin{tabular}{l c c c c c c c}

\specialrule{.1em}{0em}{0em} 
Model               & BLEU-1 & BLEU-2 & BLEU-3 & BLEU-4 & ROUGE-L & METEOR & Coverage \\ \midrule
Q-GPT-0             & 4.31   & 2.49   & 1.79   & 1.45   & 5.21    & 5.60   & 1.33     \\ 
QF-GPT-0            & 5.49   & 3.21   & 2.30   & 1.84   & 7.50    & 8.32   & 7.53     \\ 
Prompt-0            & 5.25   & 3.00   & 2.13   & 1.70   & 6.42    & 7.38   & 7.89     \\ 
Template-0          & 21.27  & 15.26  & 11.99  & 9.96   & 27.58   & 24.04  & 6.86    \\ \midrule
Subject-constrained & 29.16  & 20.99  & 16.52  & 13.68  & 28.59   & 34.78  & 9.82     \\ \midrule
QF-GPT              & 27.75  & 19.27  & 14.86  & 12.10  & 28.56   & 31.71  & 20.85    \\ 
Prompt finetuning   & 32.79  & 23.69  & 18.65  & 15.39  & 37.55   & 40.81  & 72.55    \\ 
Template-facet      & 38.56  & 26.32  & 20.87  & 17.61  & 33.87   & 46.02  & 100.0*    \\ \midrule
ZSFC + PPL          & 42.78  & 30.90  & 24.40  & 20.09  & 40.98   & 45.16  & 97.82    \\ 
ZSFC + AutoScore    & \textbf{43.42}  & \textbf{31.95}  & \textbf{25.61}  & \textbf{21.61}   & \textbf{41.03}   & \textbf{47.87}  & 98.28    \\ 
ZSFC + Cross        & 41.37  & 29.51  & 23.11  & 19.06  & 39.78   & 44.18  & 91.65    \\ 
ZSFC + NTES         & 36.92  & 26.71  & 21.13  & 17.70  & 35.69   & 41.49  & 77.32    \\ 
ZSFC + WSDM         & 41.80  & 29.69  & 23.03  & 18.88  & 38.52   & 44.19  & \textbf{98.96}    \\ 
\specialrule{.1em}{0em}{0em} 
\end{tabular}
\vspace{-5pt}
\label{table:fullquestion}
\end{table*}

\begin{table*}[t]
\caption{The effectiveness of question body only evaluation. The first number is the metric scores on question body. The second number ($\Delta$) is the performance gap between question-body and full-question evaluation, which indicates the boosted portion by question templates.}
\vspace{-7pt}
\begin{tabular}{l c c c c c c c}

\specialrule{.1em}{0em}{0em} 
Model               & BLEU-1 ($\Delta$)& BLEU-2 ($\Delta$)& BLEU-3 ($\Delta$)& BLEU-4 ($\Delta$)& ROUGE-L ($\Delta$)& METEOR ($\Delta$)& Coverage ($\Delta$)\\ \midrule
Q-GPT-0             & 2.32 (-1.99)  & 1.37 (-1.12)   & 1.03 (-0.76)  & 0.87 (-0.58)  & 3.31 (-1.90)  & 3.71 (-1.89) &  0.92 (-0.41)    \\ 
QF-GPT-0            & 3.54 (-1.95)   & 2.14 (-1.07)   & 1.56 (-0.74)  & 1.27 (-0.57)  & 5.85 (-1.65)  & 6.95 (-1.37) &  6.22 (-1.31)   \\ 
Prompt-0            & 3.40 (-1.85)  & 2.07  (-0.93)  & 1.51 (-0.62)  & 1.23 (-0.47)  & 4.52 (-1.90)  & 5.58 (-1.80) &  6.04 (-1.85)   \\ 
Template-0          & 7.89 (-13.38) & 5.22 (-10.04)  &  3.89 (-8.10) &  3.21 (-6.75)  & 10.65 (-16.93)  & 11.13 (-12.91)  & 6.27 (-0.59)    \\\midrule
Subject-constrained & 14.51 (-14.65)  & 9.44 (-11.55) & 6.97 (-9.55)  & 5.87 (-7.81)  & 14.51 (-14.08)  & 19.39 (-15.39)  & 9.45 (-0.37) \\ \midrule
QF-GPT              & 16.38 (-11.37)  & 10.92 (-8.35)  & 8.17 (-6.69)  & 6.72 (-5.38)  & 18.41 (-10.15)  & 20.14 (-11.03) & 20.21 (-0.64)    \\ 
Prompt finetuning   & 24.13 (-8.66)  & 17.25 (-6.44)  & 13.41 (-5.24)  & 11.18 (-4.21)  & 31.84 (-5.71)  & 32.99 (-7.82) & 71.41 (-1.14)   \\  
Template-facet      & 25.49 (-13.07)  & 15.62 (-10.7)  & 11.50 (-9.37)  & 9.47 (-8.14)  & 22.09 (-11.78)  & 38.58 (-7.44)  & 100.0 (0)*   \\  \midrule
ZSFC + PPL          & 36.80 (-5.98)  & 26.71 (-4.19) & 20.85 (-3.35)  & 17.26 (-2.83)  & \textbf{37.77} (-3.21)  & 40.58 (-4.58) & 97.47 (-0.35) \\ 
ZSFC + AutoScore    & 34.88 (-8.54) & 25.57 (-6.38) & 20.53 (-5.08)  & 17.42 (-4.19)  & 33.88 (-7.15)  & 43.42 (-4.45)  & 95.83 (-2.45)   \\ 
ZSFC + Cross        & 36.25 (-5.12) & 26.57 (-2.94) & 21.04 (-2.07)  & 17.59 (-1.47)  & 37.25 (-2.53)  & 40.45 (-3.73) & 90.87 (-0.78)   \\ 
ZSFC + NTES         & 29.46 (-7.46) & 21.75 (-4.96) & 17.56 (-3.57)  & 14.95 (-2.75)  & 27.97 (-7.72)  & 33.82 (-7.67) & 77.32 (\textbf{0.00})  \\ 
ZSFC + WSDM         & \textbf{38.51} (\textbf{-3.29})  & \textbf{28.34} (\textbf{-1.35})  & \textbf{22.50} (\textbf{-0.53})  & \textbf{18.99} (\textbf{+0.11})   & 37.47 (\textbf{-1.05})  & \textbf{43.79} (\textbf{-0.40})  & \textbf{98.67} (-0.29)   \\
\specialrule{.1em}{0em}{0em} 
\end{tabular}
\vspace{-5pt}

\label{table:questionbody}
\end{table*}

\subsection{Implementation Details}
We use NeuroLogic Decoding algorithm
from the author's GitHub implementation\footnote{\url{https://github.com/GXimingLu/neurologic_decoding}}. 
We implement QF-GPT by ourselves with pre-trained GPT-2 checkpoints from Huggingface and achieve similar performances as the original work \cite{sekulic2021towards}. 
Similarly, we implement prompt finetuning by changing the input of QF-GPT.

The perplexity ranker is implemented using the Huggingface pre-trained GPT-2 checkpoint for our question candidate rankers. 
We use ParlAI implementation\footnote{\url{https://github.com/facebookresearch/ParlAI}} for cross-encoder, and the ConvAI3 winning team's implementation\footnote{\url{https://github.com/ouwenjie03/Clariq_System}} for their ranker named NTES.
We implement our own Weighted Sequential Dependency Model using the intuitively adjusted parameters: $\lambda_t=\lambda_o=\lambda_u = 1, \mu=25$.

\begin{table*}[t]
\caption{Human evaluations for models in RQ.3 according to major vote from 5 annotators. Our human evaluation results are aligned with our automatic evaluations. $\dag$ and $\ddag$ indicates $p < 0.05$ and $p < 0.0001$ statistical significance over other models.}
\vspace{-7pt}
\begin{tabular}{l c c c c c c c}

\specialrule{.1em}{0em}{0em} 

\multirow{2}{*}{model} & \multicolumn{3}{c}{Naturalness} &  & \multicolumn{3}{c}{Usefulness} \\ \cline{2-4} \cline{6-8} 
                    & Good    & Fair    & Bad     & &  Good   & Fair   & Bad      \\ \midrule
QF-GPT              & 59.5\%  & 14.4\%  & 26.1\%  & & 11.5\%  & 17.4\% & 71.1\% \\ 
Prompt finetuning   & 43.7\%  & 11.1\%  & 45.2\%  & & 29.4\%  & 22.4\% & 48.2\% \\ 
Template-facet      & 57.4\%  & 32.9\%  & 9.6\%   & & 50.8\% & 36.0\%  & 13.2\% \\ 
ZSFC + WSDM         & \textbf{82.6\%}$^{\ddag}$   & 13.9\%  & \textbf{3.5\%}$^{\ddag}$ & & \textbf{68.9\%}$^{\ddag}$ & 21.9\% & \textbf{9.2\%}$^{\dag}$  \\ 
\specialrule{.1em}{0em}{0em} 
\end{tabular}
\vspace{-10pt}
\label{table:human}
\end{table*}

\section{Results and Analyses}
\label{sec:results}

This section answers the research questions using our experiment results. 
In Table~\ref{table:fullquestion} and Table~\ref{table:questionbody}, we show the automatic metrics evaluation results respectively for the full-question evaluation and the question-body evaluation, as described in Section~\ref{sec:evaluation}. 
In Table~\ref{table:human}, we show the human annotation results of the compared models in the third research question.

\footnotetext{The 100\% coverage from the Template-facet method is not included in the coverage comparison.}

Our first research question RQ1 is about the performance of existing methods for zero-shot clarifying question generation.
The results are shown in the first four rows in the both tables.
From the full question evaluation in table~\ref{table:fullquestion}, we see that all these baselines struggle to produce any reasonable generations except for Template-0.
However, we cannot conclude that Template-0 generates significantly better questions.
Because when we compare the question body evaluation results in Table~\ref{table:questionbody}, we see the scores of Template-0 drop significantly.
This means that its question body is not good, which implies that the reason for its higher score on full question evaluation is because of the question templates.
In general, we find existing zero-shot GPT-2-based approaches cannot solve the clarifying question generation task effectively.

Our second research question RQ2 is about the effectiveness of facet information for facet-specific clarifying question generation. 
To answer this question, we compare our proposed zero-shot facet-constrained (ZSFC) methods with a facet-free variation of ZSFC named Subject-constrained which uses subject of the query as constraints.
It would be unfair to compare the coverage metric between the two models because the Subject-constrained system does not access facet information. 
The gold references used for computing other metrics are the facet-specific clarifying questions from the dataset, thus it would also be incomprehensive to see the scores as the quality of clarifying question generation in general. 
Because the generated question could be reflecting another facet and get a low score on these metrics. 
However, these metrics could still be seen as the generation quality about these specific facets, which should intuitively be improved by adding facet as input, although not with naive GPT-2 \cite{sekulic2021towards}.

From both the entire generation and question body evaluations, we see that all the ZSFC models significantly improve the Subject-constrained method across all the other evaluation metrics.
The ZSFC models also drop less performance when switched to question-body evaluation, which suggests its better performance is more from the question body.
In contrast to existing works, our study show that adequate use of facet information can significantly improve clarifying question generation quality. 

\balance
The last research question RQ3 is whether our proposed zero-shot approach can perform the same or even better than existing facet-driven baselines.
To answer this question, we compare our method with a simple clarifying question rewriting baseline and two finetuning baselines in the third section of the table.
Among them, QF-GPT is the existing method \cite{sekulic2021towards}, and Prompt finetuning is our proposed prompt-based finetuning method.
We see that from both tables, our zero-shot facet-driven approaches are always better than the finetuning baselines. 
Our best-performing generation system, ZSFC+WSDM, improves the existing method QF-GPT by a large margin in the full-question evaluation and doubles its performance in question-body evaluation.
When compared with the Template-facet baseline, ZSFC can outperform it in both tables.
This implies these ZSFC system generations have more word overlaps from the reference beyond just the facet, which means the performance improvements are non-trivial.
We also bold the smallest performance drop between the two evaluations. We can see that our ZSFC models have relatively minor performance drops, which means that they potentially generate better question bodies.  

\balance
To validate the above conclusions, we employ human annotators to label the quality of generated questions. 
From Table~\ref{table:human}, we can see that our proposed system ZSFC gives the best generation for both naturalness and usefulness. 
Specifically, it generates the most amount of good naturalness and usefulness questions and the least bad ones.
This human evaluation result strengthens the conclusion from the automatic evaluation that our method is better than the baseline and supervised learning methods.
We also notice that the Template-facet rewriting is a simple yet strong baseline that both finetuning-based methods are actually worse than it. 
However, ZSFC outperforms it by a large margin in both measures.

\subsection{Ablation Study} 

We are also interested in whether question prompting is necessary for our system and which question ranker is the best in Section~\ref{sec:ranking}. 
To answer this question, we compare all five ranking methods mentioned in Section~\ref{sec:ranking}. 
For each (query, facet) pair in our dataset and each ranker, we will run the ranker on the eight question candidates using eight question templates and choose the top question from the ranked lists as the question generation of that ranker. 
The no-prompt-no-ranker approach will generate only one sentence with NeuroLogic Decoding with query as input and facet as constraints.

Here, we analyze the results of all the ranker variations in the last section from Table~\ref{table:fullquestion} and \ref{table:questionbody}.
The full-question evaluation results show that the AutoScore ranker performs best on all but the coverage metric.
However, the question-body evaluation results suggest that the WSDM ranker performs the best. 
On the one hand, we believe the question-body evaluation results are more convincing by our previous analyses and examples of the two evaluation methods.
On the other, we notice that when switched to question-body evaluation, the performances of AutoScore drop more than other methods, while WSDM almost does not change or even increase its scores. 
This could further suggest that part of the performances of AutoScore should be attributed to the question templates.
We now explain why WSDM metrics have such low performance drops. 
Unlike other ranking methods, the way the WSDM scoring function is defined encourages generation to score higher with a high-quality question body since the question templates rarely contain facet or query subject words.
Based on all the above observations and reasoning, we propose that WSDM is the best ranker.

\section{Conclusion}
In this work, we study the task of zero-shot clarifying question generation for conversational search.
We propose to solve the task as a constrained language generation problem and present a concrete system.
To demonstrate the power of our system, we answer three research questions, including comparing our zero-shot system with baseline and existing supervised learning approaches.
All the experiment results have been evaluated using a variety of natural language generation metrics, and human evaluations are done for part of the results.
The automatic metrics and human annotation results suggest our proposed zero-shot system outperforms the other compared approaches.  
Our work can be seen as both a solid zero-shot solution to the cold start problem of conversation search and a compelling demonstration of how large deep models benefit from properly integrating human knowledge.






\bibliographystyle{ACM-Reference-Format}
\bibliography{sample-sigconf}


\begin{thebibliography}{58}


\ifx \showCODEN    \undefined \def \showCODEN     #1{\unskip}     \fi
\ifx \showDOI      \undefined \def \showDOI       #1{#1}\fi
\ifx \showISBNx    \undefined \def \showISBNx     #1{\unskip}     \fi
\ifx \showISBNxiii \undefined \def \showISBNxiii  #1{\unskip}     \fi
\ifx \showISSN     \undefined \def \showISSN      #1{\unskip}     \fi
\ifx \showLCCN     \undefined \def \showLCCN      #1{\unskip}     \fi
\ifx \shownote     \undefined \def \shownote      #1{#1}          \fi
\ifx \showarticletitle \undefined \def \showarticletitle #1{#1}   \fi
\ifx \showURL      \undefined \def \showURL       {\relax}        \fi
\providecommand\bibfield[2]{#2}
\providecommand\bibinfo[2]{#2}
\providecommand\natexlab[1]{#1}
\providecommand\showeprint[2][]{arXiv:#2}

\bibitem[\protect\citeauthoryear{Aliannejadi, Kiseleva, Chuklin, Dalton, and
  Burtsev}{Aliannejadi et~al\mbox{.}}{2020}]%
        {aliannejadi2020convai3}
\bibfield{author}{\bibinfo{person}{Mohammad Aliannejadi},
  \bibinfo{person}{Julia Kiseleva}, \bibinfo{person}{Aleksandr Chuklin},
  \bibinfo{person}{Jeff Dalton}, {and} \bibinfo{person}{Mikhail Burtsev}.}
  \bibinfo{year}{2020}\natexlab{}.
\newblock \showarticletitle{ConvAI3: Generating clarifying questions for
  open-domain dialogue systems (ClariQ)}.
\newblock \bibinfo{journal}{\emph{arXiv preprint arXiv:2009.11352}}
  (\bibinfo{year}{2020}).
\newblock


\bibitem[\protect\citeauthoryear{Aliannejadi, Zamani, Crestani, and
  Croft}{Aliannejadi et~al\mbox{.}}{2019}]%
        {aliannejadi2019asking}
\bibfield{author}{\bibinfo{person}{Mohammad Aliannejadi},
  \bibinfo{person}{Hamed Zamani}, \bibinfo{person}{Fabio Crestani}, {and}
  \bibinfo{person}{W~Bruce Croft}.} \bibinfo{year}{2019}\natexlab{}.
\newblock \showarticletitle{Asking clarifying questions in open-domain
  information-seeking conversations}. In \bibinfo{booktitle}{\emph{Proceedings
  of the 42nd international acm sigir conference on research and development in
  information retrieval}}. \bibinfo{pages}{475--484}.
\newblock


\bibitem[\protect\citeauthoryear{Anand, Cavedon, Hagen, Joho, Sanderson, and
  Stein}{Anand et~al\mbox{.}}{2021}]%
        {anand2020conversational}
\bibfield{author}{\bibinfo{person}{Avishek Anand}, \bibinfo{person}{Lawrence
  Cavedon}, \bibinfo{person}{Matthias Hagen}, \bibinfo{person}{Hideo Joho},
  \bibinfo{person}{Mark Sanderson}, {and} \bibinfo{person}{Benno Stein}.}
  \bibinfo{year}{2021}\natexlab{}.
\newblock \showarticletitle{Dagstuhl Seminar 19461 on Conversational Search:
  Seminar Goals and Working Group Outcomes}.
\newblock \bibinfo{journal}{\emph{SIGIR Forum}} \bibinfo{volume}{54},
  \bibinfo{number}{1}, Article \bibinfo{articleno}{3} (\bibinfo{date}{feb}
  \bibinfo{year}{2021}), \bibinfo{numpages}{11}~pages.
\newblock
\showISSN{0163-5840}
\urldef\tempurl%
\url{https://doi.org/10.1145/3451964.3451967}
\showDOI{\tempurl}


\bibitem[\protect\citeauthoryear{Anderson, Fernando, Johnson, and
  Gould}{Anderson et~al\mbox{.}}{2017}]%
        {anderson-etal-2017-guided}
\bibfield{author}{\bibinfo{person}{Peter Anderson}, \bibinfo{person}{Basura
  Fernando}, \bibinfo{person}{Mark Johnson}, {and} \bibinfo{person}{Stephen
  Gould}.} \bibinfo{year}{2017}\natexlab{}.
\newblock \showarticletitle{Guided Open Vocabulary Image Captioning with
  Constrained Beam Search}. In \bibinfo{booktitle}{\emph{Proceedings of the
  2017 Conference on Empirical Methods in Natural Language Processing}}.
  \bibinfo{publisher}{Association for Computational Linguistics},
  \bibinfo{address}{Copenhagen, Denmark}, \bibinfo{pages}{936--945}.
\newblock
\urldef\tempurl%
\url{https://doi.org/10.18653/v1/D17-1098}
\showDOI{\tempurl}


\bibitem[\protect\citeauthoryear{Banerjee and Lavie}{Banerjee and
  Lavie}{2005}]%
        {banerjee-lavie-2005-meteor}
\bibfield{author}{\bibinfo{person}{Satanjeev Banerjee} {and}
  \bibinfo{person}{Alon Lavie}.} \bibinfo{year}{2005}\natexlab{}.
\newblock \showarticletitle{{METEOR}: An Automatic Metric for {MT} Evaluation
  with Improved Correlation with Human Judgments}. In
  \bibinfo{booktitle}{\emph{Proceedings of the {ACL} Workshop on Intrinsic and
  Extrinsic Evaluation Measures for Machine Translation and/or Summarization}}.
  \bibinfo{publisher}{Association for Computational Linguistics},
  \bibinfo{address}{Ann Arbor, Michigan}, \bibinfo{pages}{65--72}.
\newblock
\urldef\tempurl%
\url{https://aclanthology.org/W05-0909}
\showURL{%
\tempurl}


\bibitem[\protect\citeauthoryear{Belz and Reiter}{Belz and Reiter}{2006}]%
        {belz2006comparing}
\bibfield{author}{\bibinfo{person}{Anja Belz} {and} \bibinfo{person}{Ehud
  Reiter}.} \bibinfo{year}{2006}\natexlab{}.
\newblock \showarticletitle{Comparing automatic and human evaluation of NLG
  systems}. In \bibinfo{booktitle}{\emph{11th conference of the european
  chapter of the association for computational linguistics}}.
  \bibinfo{pages}{313--320}.
\newblock


\bibitem[\protect\citeauthoryear{Bendersky, Metzler, and Croft}{Bendersky
  et~al\mbox{.}}{2010}]%
        {bendersky2010learning}
\bibfield{author}{\bibinfo{person}{Michael Bendersky}, \bibinfo{person}{Donald
  Metzler}, {and} \bibinfo{person}{W.~Bruce Croft}.}
  \bibinfo{year}{2010}\natexlab{}.
\newblock \showarticletitle{Learning Concept Importance Using a Weighted
  Dependence Model}. In \bibinfo{booktitle}{\emph{Proceedings of the Third ACM
  International Conference on Web Search and Data Mining}} (New York, New York,
  USA) \emph{(\bibinfo{series}{WSDM '10})}. \bibinfo{publisher}{Association for
  Computing Machinery}, \bibinfo{address}{New York, NY, USA},
  \bibinfo{pages}{31–40}.
\newblock
\showISBNx{9781605588896}
\urldef\tempurl%
\url{https://doi.org/10.1145/1718487.1718492}
\showDOI{\tempurl}


\bibitem[\protect\citeauthoryear{Bi, Ai, and Croft}{Bi et~al\mbox{.}}{2021}]%
        {bi2021asking}
\bibfield{author}{\bibinfo{person}{Keping Bi}, \bibinfo{person}{Qingyao Ai},
  {and} \bibinfo{person}{W~Bruce Croft}.} \bibinfo{year}{2021}\natexlab{}.
\newblock \showarticletitle{Asking Clarifying Questions Based on Negative
  Feedback in Conversational Search}. In \bibinfo{booktitle}{\emph{Proceedings
  of the 2021 ACM SIGIR International Conference on Theory of Information
  Retrieval}}. \bibinfo{pages}{157--166}.
\newblock


\bibitem[\protect\citeauthoryear{Clark, Luong, Le, and Manning}{Clark
  et~al\mbox{.}}{2020}]%
        {clark2020electra}
\bibfield{author}{\bibinfo{person}{Kevin Clark}, \bibinfo{person}{Minh-Thang
  Luong}, \bibinfo{person}{Quoc~V Le}, {and} \bibinfo{person}{Christopher~D
  Manning}.} \bibinfo{year}{2020}\natexlab{}.
\newblock \showarticletitle{Electra: Pre-training text encoders as
  discriminators rather than generators}.
\newblock \bibinfo{journal}{\emph{arXiv preprint arXiv:2003.10555}}
  (\bibinfo{year}{2020}).
\newblock


\bibitem[\protect\citeauthoryear{Clarke, Craswell, and Soboroff}{Clarke
  et~al\mbox{.}}{2009}]%
        {Clarke2009OverviewOT}
\bibfield{author}{\bibinfo{person}{Charles L.~A. Clarke}, \bibinfo{person}{Nick
  Craswell}, {and} \bibinfo{person}{Ian Soboroff}.}
  \bibinfo{year}{2009}\natexlab{}.
\newblock \showarticletitle{Overview of the TREC 2009 Web Track}. In
  \bibinfo{booktitle}{\emph{TREC}}.
\newblock


\bibitem[\protect\citeauthoryear{Culpepper, Diaz, and Smucker}{Culpepper
  et~al\mbox{.}}{2018}]%
        {culpepper2018research}
\bibfield{author}{\bibinfo{person}{J~Shane Culpepper},
  \bibinfo{person}{Fernando Diaz}, {and} \bibinfo{person}{Mark~D Smucker}.}
  \bibinfo{year}{2018}\natexlab{}.
\newblock \showarticletitle{Research frontiers in information retrieval: Report
  from the third strategic workshop on information retrieval in lorne (swirl
  2018)}. In \bibinfo{booktitle}{\emph{ACM SIGIR Forum}},
  Vol.~\bibinfo{volume}{52}. ACM New York, NY, USA, \bibinfo{pages}{34--90}.
\newblock


\bibitem[\protect\citeauthoryear{Dang and Croft}{Dang and Croft}{2010}]%
        {dang2010query}
\bibfield{author}{\bibinfo{person}{Van Dang} {and} \bibinfo{person}{Bruce~W
  Croft}.} \bibinfo{year}{2010}\natexlab{}.
\newblock \showarticletitle{Query reformulation using anchor text}. In
  \bibinfo{booktitle}{\emph{Proceedings of the third ACM international
  conference on Web search and data mining}}. \bibinfo{pages}{41--50}.
\newblock


\bibitem[\protect\citeauthoryear{Dhole}{Dhole}{2020}]%
        {dhole2020resolving}
\bibfield{author}{\bibinfo{person}{Kaustubh~D Dhole}.}
  \bibinfo{year}{2020}\natexlab{}.
\newblock \showarticletitle{Resolving intent ambiguities by retrieving
  discriminative clarifying questions}.
\newblock \bibinfo{journal}{\emph{arXiv preprint arXiv:2008.07559}}
  (\bibinfo{year}{2020}).
\newblock


\bibitem[\protect\citeauthoryear{Elgohary, Peskov, and Boyd-Graber}{Elgohary
  et~al\mbox{.}}{2019}]%
        {elgohary2019can}
\bibfield{author}{\bibinfo{person}{Ahmed Elgohary}, \bibinfo{person}{Denis
  Peskov}, {and} \bibinfo{person}{Jordan Boyd-Graber}.}
  \bibinfo{year}{2019}\natexlab{}.
\newblock \showarticletitle{Can you unpack that? learning to rewrite
  questions-in-context}.
\newblock \bibinfo{journal}{\emph{Can You Unpack That? Learning to Rewrite
  Questions-in-Context}} (\bibinfo{year}{2019}).
\newblock


\bibitem[\protect\citeauthoryear{Fu, Xian, Zhang, and Zhang}{Fu
  et~al\mbox{.}}{2020}]%
        {fu2020tutorial}
\bibfield{author}{\bibinfo{person}{Zuohui Fu}, \bibinfo{person}{Yikun Xian},
  \bibinfo{person}{Yongfeng Zhang}, {and} \bibinfo{person}{Yi Zhang}.}
  \bibinfo{year}{2020}\natexlab{}.
\newblock \showarticletitle{Tutorial on Conversational Recommendation Systems}.
  In \bibinfo{booktitle}{\emph{Fourteenth ACM Conference on Recommender
  Systems}}. \bibinfo{pages}{751--753}.
\newblock


\bibitem[\protect\citeauthoryear{Gao, Galley, and Li}{Gao
  et~al\mbox{.}}{2018}]%
        {gao2018neural}
\bibfield{author}{\bibinfo{person}{Jianfeng Gao}, \bibinfo{person}{Michel
  Galley}, {and} \bibinfo{person}{Lihong Li}.} \bibinfo{year}{2018}\natexlab{}.
\newblock \showarticletitle{Neural approaches to conversational ai}. In
  \bibinfo{booktitle}{\emph{The 41st International ACM SIGIR Conference on
  Research \& Development in Information Retrieval}}.
  \bibinfo{pages}{1371--1374}.
\newblock


\bibitem[\protect\citeauthoryear{Gao, Xiong, Bennett, and Craswell}{Gao
  et~al\mbox{.}}{2022}]%
        {gao2022neural}
\bibfield{author}{\bibinfo{person}{Jianfeng Gao}, \bibinfo{person}{Chenyan
  Xiong}, \bibinfo{person}{Paul Bennett}, {and} \bibinfo{person}{Nick
  Craswell}.} \bibinfo{year}{2022}\natexlab{}.
\newblock \showarticletitle{Neural approaches to conversational information
  retrieval}.
\newblock \bibinfo{journal}{\emph{arXiv preprint arXiv:2201.05176}}
  (\bibinfo{year}{2022}).
\newblock


\bibitem[\protect\citeauthoryear{Hauff, Kiseleva, Sanderson, Zamani, and
  Zhang}{Hauff et~al\mbox{.}}{2021}]%
        {hauff2021conversational}
\bibfield{author}{\bibinfo{person}{Claudia Hauff}, \bibinfo{person}{Julia
  Kiseleva}, \bibinfo{person}{Mark Sanderson}, \bibinfo{person}{Hamed Zamani},
  {and} \bibinfo{person}{Yongfeng Zhang}.} \bibinfo{year}{2021}\natexlab{}.
\newblock \bibinfo{title}{Conversational Search and Recommendation:
  Introduction to the Special Issue}.
\newblock
\newblock


\bibitem[\protect\citeauthoryear{Hokamp and Liu}{Hokamp and Liu}{2017}]%
        {hokamp-liu-2017-lexically}
\bibfield{author}{\bibinfo{person}{Chris Hokamp} {and} \bibinfo{person}{Qun
  Liu}.} \bibinfo{year}{2017}\natexlab{}.
\newblock \showarticletitle{Lexically Constrained Decoding for Sequence
  Generation Using Grid Beam Search}. In \bibinfo{booktitle}{\emph{Proceedings
  of the 55th Annual Meeting of the Association for Computational Linguistics
  (Volume 1: Long Papers)}}. \bibinfo{publisher}{Association for Computational
  Linguistics}, \bibinfo{address}{Vancouver, Canada},
  \bibinfo{pages}{1535--1546}.
\newblock
\urldef\tempurl%
\url{https://doi.org/10.18653/v1/P17-1141}
\showDOI{\tempurl}


\bibitem[\protect\citeauthoryear{Humeau, Shuster, Lachaux, and Weston}{Humeau
  et~al\mbox{.}}{2019}]%
        {humeau2019poly}
\bibfield{author}{\bibinfo{person}{Samuel Humeau}, \bibinfo{person}{Kurt
  Shuster}, \bibinfo{person}{Marie-Anne Lachaux}, {and} \bibinfo{person}{Jason
  Weston}.} \bibinfo{year}{2019}\natexlab{}.
\newblock \showarticletitle{Poly-encoders: Transformer architectures and
  pre-training strategies for fast and accurate multi-sentence scoring}.
\newblock \bibinfo{journal}{\emph{arXiv preprint arXiv:1905.01969}}
  (\bibinfo{year}{2019}).
\newblock


\bibitem[\protect\citeauthoryear{Keyvan and Huang}{Keyvan and Huang}{2022}]%
        {keyvan2022approach}
\bibfield{author}{\bibinfo{person}{Kimiya Keyvan} {and}
  \bibinfo{person}{Jimmy~Xiangji Huang}.} \bibinfo{year}{2022}\natexlab{}.
\newblock \showarticletitle{How to Approach Ambiguous Queries in Conversational
  Search? A Survey of Techniques, Approaches, Tools and Challenges}.
\newblock \bibinfo{journal}{\emph{ACM Computing Surveys (CSUR)}}
  (\bibinfo{year}{2022}).
\newblock


\bibitem[\protect\citeauthoryear{Krasakis, Aliannejadi, Voskarides, and
  Kanoulas}{Krasakis et~al\mbox{.}}{2020}]%
        {krasakis2020analysing}
\bibfield{author}{\bibinfo{person}{Antonios~Minas Krasakis},
  \bibinfo{person}{Mohammad Aliannejadi}, \bibinfo{person}{Nikos Voskarides},
  {and} \bibinfo{person}{Evangelos Kanoulas}.} \bibinfo{year}{2020}\natexlab{}.
\newblock \showarticletitle{Analysing the effect of clarifying questions on
  document ranking in conversational search}. In
  \bibinfo{booktitle}{\emph{Proceedings of the 2020 acm sigir on international
  conference on theory of information retrieval}}. \bibinfo{pages}{129--132}.
\newblock


\bibitem[\protect\citeauthoryear{Li and Liang}{Li and Liang}{2021}]%
        {li2021prefix}
\bibfield{author}{\bibinfo{person}{Xiang~Lisa Li} {and} \bibinfo{person}{Percy
  Liang}.} \bibinfo{year}{2021}\natexlab{}.
\newblock \showarticletitle{Prefix-tuning: Optimizing continuous prompts for
  generation}.
\newblock \bibinfo{journal}{\emph{arXiv preprint arXiv:2101.00190}}
  (\bibinfo{year}{2021}).
\newblock


\bibitem[\protect\citeauthoryear{Lin, Zhou, Shen, Zhou, Bhagavatula, Choi, and
  Ren}{Lin et~al\mbox{.}}{2020}]%
        {lin-etal-2020-commongen}
\bibfield{author}{\bibinfo{person}{Bill~Yuchen Lin},
  \bibinfo{person}{Wangchunshu Zhou}, \bibinfo{person}{Ming Shen},
  \bibinfo{person}{Pei Zhou}, \bibinfo{person}{Chandra Bhagavatula},
  \bibinfo{person}{Yejin Choi}, {and} \bibinfo{person}{Xiang Ren}.}
  \bibinfo{year}{2020}\natexlab{}.
\newblock \showarticletitle{{C}ommon{G}en: A Constrained Text Generation
  Challenge for Generative Commonsense Reasoning}. In
  \bibinfo{booktitle}{\emph{Findings of the Association for Computational
  Linguistics: EMNLP 2020}}. \bibinfo{publisher}{Association for Computational
  Linguistics}, \bibinfo{address}{Online}, \bibinfo{pages}{1823--1840}.
\newblock
\urldef\tempurl%
\url{https://doi.org/10.18653/v1/2020.findings-emnlp.165}
\showDOI{\tempurl}


\bibitem[\protect\citeauthoryear{Lin}{Lin}{2004}]%
        {lin-2004-rouge}
\bibfield{author}{\bibinfo{person}{Chin-Yew Lin}.}
  \bibinfo{year}{2004}\natexlab{}.
\newblock \showarticletitle{{ROUGE}: A Package for Automatic Evaluation of
  Summaries}. In \bibinfo{booktitle}{\emph{Text Summarization Branches Out}}.
  \bibinfo{publisher}{Association for Computational Linguistics},
  \bibinfo{address}{Barcelona, Spain}, \bibinfo{pages}{74--81}.
\newblock
\urldef\tempurl%
\url{https://aclanthology.org/W04-1013}
\showURL{%
\tempurl}


\bibitem[\protect\citeauthoryear{Liu, Gwizdka, Liu, Xu, and Belkin}{Liu
  et~al\mbox{.}}{2010}]%
        {liu2010analysis}
\bibfield{author}{\bibinfo{person}{Chang Liu}, \bibinfo{person}{Jacek Gwizdka},
  \bibinfo{person}{Jingjing Liu}, \bibinfo{person}{Tao Xu}, {and}
  \bibinfo{person}{Nicholas~J Belkin}.} \bibinfo{year}{2010}\natexlab{}.
\newblock \showarticletitle{Analysis and evaluation of query reformulations in
  different task types}.
\newblock \bibinfo{journal}{\emph{Proceedings of the American Society for
  Information Science and Technology}} \bibinfo{volume}{47},
  \bibinfo{number}{1} (\bibinfo{year}{2010}), \bibinfo{pages}{1--9}.
\newblock


\bibitem[\protect\citeauthoryear{Liu, Yuan, Fu, Jiang, Hayashi, and Neubig}{Liu
  et~al\mbox{.}}{2021}]%
        {liu2021pre}
\bibfield{author}{\bibinfo{person}{Pengfei Liu}, \bibinfo{person}{Weizhe Yuan},
  \bibinfo{person}{Jinlan Fu}, \bibinfo{person}{Zhengbao Jiang},
  \bibinfo{person}{Hiroaki Hayashi}, {and} \bibinfo{person}{Graham Neubig}.}
  \bibinfo{year}{2021}\natexlab{}.
\newblock \showarticletitle{Pre-train, prompt, and predict: A systematic survey
  of prompting methods in natural language processing}.
\newblock \bibinfo{journal}{\emph{arXiv preprint arXiv:2107.13586}}
  (\bibinfo{year}{2021}).
\newblock


\bibitem[\protect\citeauthoryear{Lu, Welleck, West, Jiang, Kasai, Khashabi,
  Bras, Qin, Yu, Zellers, et~al\mbox{.}}{Lu et~al\mbox{.}}{2021}]%
        {lu2021neurologic}
\bibfield{author}{\bibinfo{person}{Ximing Lu}, \bibinfo{person}{Sean Welleck},
  \bibinfo{person}{Peter West}, \bibinfo{person}{Liwei Jiang},
  \bibinfo{person}{Jungo Kasai}, \bibinfo{person}{Daniel Khashabi},
  \bibinfo{person}{Ronan~Le Bras}, \bibinfo{person}{Lianhui Qin},
  \bibinfo{person}{Youngjae Yu}, \bibinfo{person}{Rowan Zellers},
  {et~al\mbox{.}}} \bibinfo{year}{2021}\natexlab{}.
\newblock \showarticletitle{Neurologic a* esque decoding: Constrained text
  generation with lookahead heuristics}.
\newblock \bibinfo{journal}{\emph{arXiv preprint arXiv:2112.08726}}
  (\bibinfo{year}{2021}).
\newblock


\bibitem[\protect\citeauthoryear{Lu, West, Zellers, Bras, Bhagavatula, and
  Choi}{Lu et~al\mbox{.}}{2020}]%
        {lu2020neurologic}
\bibfield{author}{\bibinfo{person}{Ximing Lu}, \bibinfo{person}{Peter West},
  \bibinfo{person}{Rowan Zellers}, \bibinfo{person}{Ronan~Le Bras},
  \bibinfo{person}{Chandra Bhagavatula}, {and} \bibinfo{person}{Yejin Choi}.}
  \bibinfo{year}{2020}\natexlab{}.
\newblock \showarticletitle{Neurologic decoding:(un) supervised neural text
  generation with predicate logic constraints}.
\newblock \bibinfo{journal}{\emph{arXiv preprint arXiv:2010.12884}}
  (\bibinfo{year}{2020}).
\newblock


\bibitem[\protect\citeauthoryear{Marchionini}{Marchionini}{2006}]%
        {marchionini2006exploratory}
\bibfield{author}{\bibinfo{person}{Gary Marchionini}.}
  \bibinfo{year}{2006}\natexlab{}.
\newblock \showarticletitle{Exploratory Search: From Finding to Understanding}.
\newblock \bibinfo{journal}{\emph{Commun. ACM}} \bibinfo{volume}{49},
  \bibinfo{number}{4} (\bibinfo{date}{apr} \bibinfo{year}{2006}),
  \bibinfo{pages}{41–46}.
\newblock
\showISSN{0001-0782}
\urldef\tempurl%
\url{https://doi.org/10.1145/1121949.1121979}
\showDOI{\tempurl}


\bibitem[\protect\citeauthoryear{Mazar{\'e}, Humeau, Raison, and
  Bordes}{Mazar{\'e} et~al\mbox{.}}{2018}]%
        {mazare2018training}
\bibfield{author}{\bibinfo{person}{Pierre-Emmanuel Mazar{\'e}},
  \bibinfo{person}{Samuel Humeau}, \bibinfo{person}{Martin Raison}, {and}
  \bibinfo{person}{Antoine Bordes}.} \bibinfo{year}{2018}\natexlab{}.
\newblock \showarticletitle{Training millions of personalized dialogue agents}.
\newblock \bibinfo{journal}{\emph{arXiv preprint arXiv:1809.01984}}
  (\bibinfo{year}{2018}).
\newblock


\bibitem[\protect\citeauthoryear{McCloskey and Cohen}{McCloskey and
  Cohen}{1989}]%
        {mccloskey1989catastrophic}
\bibfield{author}{\bibinfo{person}{Michael McCloskey} {and}
  \bibinfo{person}{Neal~J Cohen}.} \bibinfo{year}{1989}\natexlab{}.
\newblock \showarticletitle{Catastrophic interference in connectionist
  networks: The sequential learning problem}.
\newblock In \bibinfo{booktitle}{\emph{Psychology of learning and motivation}}.
  Vol.~\bibinfo{volume}{24}. \bibinfo{publisher}{Elsevier},
  \bibinfo{pages}{109--165}.
\newblock


\bibitem[\protect\citeauthoryear{Miao, Zhou, Mou, Yan, and Li}{Miao
  et~al\mbox{.}}{2019}]%
        {miao2019cgmh}
\bibfield{author}{\bibinfo{person}{Ning Miao}, \bibinfo{person}{Hao Zhou},
  \bibinfo{person}{Lili Mou}, \bibinfo{person}{Rui Yan}, {and}
  \bibinfo{person}{Lei Li}.} \bibinfo{year}{2019}\natexlab{}.
\newblock \showarticletitle{Cgmh: Constrained sentence generation by
  metropolis-hastings sampling}. In \bibinfo{booktitle}{\emph{Proceedings of
  the AAAI Conference on Artificial Intelligence}}, Vol.~\bibinfo{volume}{33}.
  \bibinfo{pages}{6834--6842}.
\newblock


\bibitem[\protect\citeauthoryear{Ou and Lin}{Ou and Lin}{2020}]%
        {ou2020clarifying}
\bibfield{author}{\bibinfo{person}{Wenjie Ou} {and} \bibinfo{person}{Yue Lin}.}
  \bibinfo{year}{2020}\natexlab{}.
\newblock \showarticletitle{A clarifying question selection system from
  ntes\_along in convai3 challenge}.
\newblock \bibinfo{journal}{\emph{arXiv preprint arXiv:2010.14202}}
  (\bibinfo{year}{2020}).
\newblock


\bibitem[\protect\citeauthoryear{Papineni, Roukos, Ward, and Zhu}{Papineni
  et~al\mbox{.}}{2002}]%
        {papineni-etal-2002-bleu}
\bibfield{author}{\bibinfo{person}{Kishore Papineni}, \bibinfo{person}{Salim
  Roukos}, \bibinfo{person}{Todd Ward}, {and} \bibinfo{person}{Wei-Jing Zhu}.}
  \bibinfo{year}{2002}\natexlab{}.
\newblock \showarticletitle{{B}leu: a Method for Automatic Evaluation of
  Machine Translation}. In \bibinfo{booktitle}{\emph{Proceedings of the 40th
  Annual Meeting of the Association for Computational Linguistics}}.
  \bibinfo{publisher}{Association for Computational Linguistics},
  \bibinfo{address}{Philadelphia, Pennsylvania, USA},
  \bibinfo{pages}{311--318}.
\newblock
\urldef\tempurl%
\url{https://doi.org/10.3115/1073083.1073135}
\showDOI{\tempurl}


\bibitem[\protect\citeauthoryear{Radford, Wu, Child, Luan, Amodei, Sutskever,
  et~al\mbox{.}}{Radford et~al\mbox{.}}{2019}]%
        {radford2019language}
\bibfield{author}{\bibinfo{person}{Alec Radford}, \bibinfo{person}{Jeffrey Wu},
  \bibinfo{person}{Rewon Child}, \bibinfo{person}{David Luan},
  \bibinfo{person}{Dario Amodei}, \bibinfo{person}{Ilya Sutskever},
  {et~al\mbox{.}}} \bibinfo{year}{2019}\natexlab{}.
\newblock \showarticletitle{Language models are unsupervised multitask
  learners}.
\newblock \bibinfo{journal}{\emph{OpenAI blog}} \bibinfo{volume}{1},
  \bibinfo{number}{8} (\bibinfo{year}{2019}), \bibinfo{pages}{9}.
\newblock


\bibitem[\protect\citeauthoryear{Radlinski and Craswell}{Radlinski and
  Craswell}{2017}]%
        {radlinski2017theoretical}
\bibfield{author}{\bibinfo{person}{Filip Radlinski} {and} \bibinfo{person}{Nick
  Craswell}.} \bibinfo{year}{2017}\natexlab{}.
\newblock \showarticletitle{A theoretical framework for conversational search}.
  In \bibinfo{booktitle}{\emph{Proceedings of the 2017 conference on conference
  human information interaction and retrieval}}. \bibinfo{pages}{117--126}.
\newblock


\bibitem[\protect\citeauthoryear{Rao and Daum{\'e}~III}{Rao and
  Daum{\'e}~III}{2018}]%
        {rao2018learning}
\bibfield{author}{\bibinfo{person}{Sudha Rao} {and} \bibinfo{person}{Hal
  Daum{\'e}~III}.} \bibinfo{year}{2018}\natexlab{}.
\newblock \showarticletitle{Learning to ask good questions: Ranking
  clarification questions using neural expected value of perfect information}.
\newblock \bibinfo{journal}{\emph{arXiv preprint arXiv:1805.04655}}
  (\bibinfo{year}{2018}).
\newblock


\bibitem[\protect\citeauthoryear{Rao and Daum{\'e}~III}{Rao and
  Daum{\'e}~III}{2019}]%
        {rao2019answer}
\bibfield{author}{\bibinfo{person}{Sudha Rao} {and} \bibinfo{person}{Hal
  Daum{\'e}~III}.} \bibinfo{year}{2019}\natexlab{}.
\newblock \showarticletitle{Answer-based adversarial training for generating
  clarification questions}.
\newblock \bibinfo{journal}{\emph{arXiv preprint arXiv:1904.02281}}
  (\bibinfo{year}{2019}).
\newblock


\bibitem[\protect\citeauthoryear{Rosset, Xiong, Song, Campos, Craswell, Tiwary,
  and Bennett}{Rosset et~al\mbox{.}}{2020}]%
        {rosset2020leading}
\bibfield{author}{\bibinfo{person}{Corbin Rosset}, \bibinfo{person}{Chenyan
  Xiong}, \bibinfo{person}{Xia Song}, \bibinfo{person}{Daniel Campos},
  \bibinfo{person}{Nick Craswell}, \bibinfo{person}{Saurabh Tiwary}, {and}
  \bibinfo{person}{Paul Bennett}.} \bibinfo{year}{2020}\natexlab{}.
\newblock \showarticletitle{Leading conversational search by suggesting useful
  questions}. In \bibinfo{booktitle}{\emph{Proceedings of The Web Conference
  2020}}. \bibinfo{pages}{1160--1170}.
\newblock


\bibitem[\protect\citeauthoryear{Santos, Macdonald, Ounis,
  et~al\mbox{.}}{Santos et~al\mbox{.}}{2015}]%
        {santos2015search}
\bibfield{author}{\bibinfo{person}{Rodrygo~LT Santos}, \bibinfo{person}{Craig
  Macdonald}, \bibinfo{person}{Iadh Ounis}, {et~al\mbox{.}}}
  \bibinfo{year}{2015}\natexlab{}.
\newblock \showarticletitle{Search result diversification}.
\newblock \bibinfo{journal}{\emph{Foundations and Trends{\textregistered} in
  Information Retrieval}} \bibinfo{volume}{9}, \bibinfo{number}{1}
  (\bibinfo{year}{2015}), \bibinfo{pages}{1--90}.
\newblock


\bibitem[\protect\citeauthoryear{Schick and Sch{\"u}tze}{Schick and
  Sch{\"u}tze}{2021}]%
        {schick2021few}
\bibfield{author}{\bibinfo{person}{Timo Schick} {and} \bibinfo{person}{Hinrich
  Sch{\"u}tze}.} \bibinfo{year}{2021}\natexlab{}.
\newblock \showarticletitle{Few-shot text generation with natural language
  instructions}. Association for Computational Linguistics.
\newblock


\bibitem[\protect\citeauthoryear{Schlangen}{Schlangen}{2004}]%
        {schlangen2004causes}
\bibfield{author}{\bibinfo{person}{David Schlangen}.}
  \bibinfo{year}{2004}\natexlab{}.
\newblock \showarticletitle{Causes and strategies for requesting clarification
  in dialogue}. In \bibinfo{booktitle}{\emph{Proceedings of the 5th SIGdial
  Workshop on Discourse and Dialogue at HLT-NAACL 2004}}.
  \bibinfo{pages}{136--143}.
\newblock


\bibitem[\protect\citeauthoryear{Sekuli{\'c}, Aliannejadi, and
  Crestani}{Sekuli{\'c} et~al\mbox{.}}{2021}]%
        {sekulic2021towards}
\bibfield{author}{\bibinfo{person}{Ivan Sekuli{\'c}}, \bibinfo{person}{Mohammad
  Aliannejadi}, {and} \bibinfo{person}{Fabio Crestani}.}
  \bibinfo{year}{2021}\natexlab{}.
\newblock \showarticletitle{Towards Facet-Driven Generation of Clarifying
  Questions for Conversational Search}. In
  \bibinfo{booktitle}{\emph{Proceedings of the 2021 ACM SIGIR International
  Conference on Theory of Information Retrieval}}. \bibinfo{pages}{167--175}.
\newblock


\bibitem[\protect\citeauthoryear{Sordoni, Bengio, Vahabi, Lioma, Grue~Simonsen,
  and Nie}{Sordoni et~al\mbox{.}}{2015}]%
        {sordoni2015hierarchical}
\bibfield{author}{\bibinfo{person}{Alessandro Sordoni}, \bibinfo{person}{Yoshua
  Bengio}, \bibinfo{person}{Hossein Vahabi}, \bibinfo{person}{Christina Lioma},
  \bibinfo{person}{Jakob Grue~Simonsen}, {and} \bibinfo{person}{Jian-Yun Nie}.}
  \bibinfo{year}{2015}\natexlab{}.
\newblock \showarticletitle{A hierarchical recurrent encoder-decoder for
  generative context-aware query suggestion}. In
  \bibinfo{booktitle}{\emph{proceedings of the 24th ACM international on
  conference on information and knowledge management}}.
  \bibinfo{pages}{553--562}.
\newblock


\bibitem[\protect\citeauthoryear{Vtyurina, Savenkov, Agichtein, and
  Clarke}{Vtyurina et~al\mbox{.}}{2017}]%
        {vtyurina2017exploring}
\bibfield{author}{\bibinfo{person}{Alexandra Vtyurina}, \bibinfo{person}{Denis
  Savenkov}, \bibinfo{person}{Eugene Agichtein}, {and}
  \bibinfo{person}{Charles~LA Clarke}.} \bibinfo{year}{2017}\natexlab{}.
\newblock \showarticletitle{Exploring conversational search with humans,
  assistants, and wizards}. In \bibinfo{booktitle}{\emph{Proceedings of the
  2017 chi conference extended abstracts on human factors in computing
  systems}}. \bibinfo{pages}{2187--2193}.
\newblock


\bibitem[\protect\citeauthoryear{Wang and Li}{Wang and Li}{2021}]%
        {wang2021template}
\bibfield{author}{\bibinfo{person}{Jian Wang} {and} \bibinfo{person}{Wenjie
  Li}.} \bibinfo{year}{2021}\natexlab{}.
\newblock \showarticletitle{Template-guided Clarifying Question Generation for
  Web Search Clarification}. In \bibinfo{booktitle}{\emph{Proceedings of the
  30th ACM International Conference on Information \& Knowledge Management}}.
  \bibinfo{pages}{3468--3472}.
\newblock


\bibitem[\protect\citeauthoryear{Welleck, Brantley, Iii, and Cho}{Welleck
  et~al\mbox{.}}{2019}]%
        {welleck2019non}
\bibfield{author}{\bibinfo{person}{Sean Welleck}, \bibinfo{person}{Kiant{\'e}
  Brantley}, \bibinfo{person}{Hal~Daum{\'e} Iii}, {and}
  \bibinfo{person}{Kyunghyun Cho}.} \bibinfo{year}{2019}\natexlab{}.
\newblock \showarticletitle{Non-monotonic sequential text generation}. In
  \bibinfo{booktitle}{\emph{International Conference on Machine Learning}}.
  PMLR, \bibinfo{pages}{6716--6726}.
\newblock


\bibitem[\protect\citeauthoryear{White, Poesia, Hawkins, Sadigh, and
  Goodman}{White et~al\mbox{.}}{2021}]%
        {white-etal-2021-open}
\bibfield{author}{\bibinfo{person}{Julia White}, \bibinfo{person}{Gabriel
  Poesia}, \bibinfo{person}{Robert Hawkins}, \bibinfo{person}{Dorsa Sadigh},
  {and} \bibinfo{person}{Noah Goodman}.} \bibinfo{year}{2021}\natexlab{}.
\newblock \showarticletitle{Open-domain clarification question generation
  without question examples}. In \bibinfo{booktitle}{\emph{Proceedings of the
  2021 Conference on Empirical Methods in Natural Language Processing}}.
  \bibinfo{publisher}{Association for Computational Linguistics},
  \bibinfo{address}{Online and Punta Cana, Dominican Republic},
  \bibinfo{pages}{563--570}.
\newblock
\urldef\tempurl%
\url{https://doi.org/10.18653/v1/2021.emnlp-main.44}
\showDOI{\tempurl}


\bibitem[\protect\citeauthoryear{Xu, Wang, Tang, Duan, Yang, Zeng, Zhou, and
  Sun}{Xu et~al\mbox{.}}{2019}]%
        {xu-etal-2019-asking}
\bibfield{author}{\bibinfo{person}{Jingjing Xu}, \bibinfo{person}{Yuechen
  Wang}, \bibinfo{person}{Duyu Tang}, \bibinfo{person}{Nan Duan},
  \bibinfo{person}{Pengcheng Yang}, \bibinfo{person}{Qi Zeng},
  \bibinfo{person}{Ming Zhou}, {and} \bibinfo{person}{Xu Sun}.}
  \bibinfo{year}{2019}\natexlab{}.
\newblock \showarticletitle{Asking Clarification Questions in Knowledge-Based
  Question Answering}. In \bibinfo{booktitle}{\emph{Proceedings of the 2019
  Conference on Empirical Methods in Natural Language Processing and the 9th
  International Joint Conference on Natural Language Processing
  (EMNLP-IJCNLP)}}. \bibinfo{publisher}{Association for Computational
  Linguistics}, \bibinfo{address}{Hong Kong, China},
  \bibinfo{pages}{1618--1629}.
\newblock
\urldef\tempurl%
\url{https://doi.org/10.18653/v1/D19-1172}
\showDOI{\tempurl}


\bibitem[\protect\citeauthoryear{Yang, Zamani, Zhang, Guo, and Croft}{Yang
  et~al\mbox{.}}{2017}]%
        {yang2017neural}
\bibfield{author}{\bibinfo{person}{Liu Yang}, \bibinfo{person}{Hamed Zamani},
  \bibinfo{person}{Yongfeng Zhang}, \bibinfo{person}{Jiafeng Guo}, {and}
  \bibinfo{person}{W~Bruce Croft}.} \bibinfo{year}{2017}\natexlab{}.
\newblock \showarticletitle{Neural matching models for question retrieval and
  next question prediction in conversation}.
\newblock \bibinfo{journal}{\emph{arXiv preprint arXiv:1707.05409}}
  (\bibinfo{year}{2017}).
\newblock


\bibitem[\protect\citeauthoryear{Yu, Liu, Yang, Xiong, Bennett, Gao, and
  Liu}{Yu et~al\mbox{.}}{2020}]%
        {yu2020few}
\bibfield{author}{\bibinfo{person}{Shi Yu}, \bibinfo{person}{Jiahua Liu},
  \bibinfo{person}{Jingqin Yang}, \bibinfo{person}{Chenyan Xiong},
  \bibinfo{person}{Paul Bennett}, \bibinfo{person}{Jianfeng Gao}, {and}
  \bibinfo{person}{Zhiyuan Liu}.} \bibinfo{year}{2020}\natexlab{}.
\newblock \showarticletitle{Few-shot generative conversational query
  rewriting}. In \bibinfo{booktitle}{\emph{Proceedings of the 43rd
  International ACM SIGIR conference on research and development in Information
  Retrieval}}. \bibinfo{pages}{1933--1936}.
\newblock


\bibitem[\protect\citeauthoryear{Zamani and Craswell}{Zamani and
  Craswell}{2020}]%
        {zamani2020macaw}
\bibfield{author}{\bibinfo{person}{Hamed Zamani} {and} \bibinfo{person}{Nick
  Craswell}.} \bibinfo{year}{2020}\natexlab{}.
\newblock \showarticletitle{Macaw: An extensible conversational information
  seeking platform}. In \bibinfo{booktitle}{\emph{Proceedings of the 43rd
  International ACM SIGIR Conference on Research and Development in Information
  Retrieval}}. \bibinfo{pages}{2193--2196}.
\newblock


\bibitem[\protect\citeauthoryear{Zamani, Dumais, Craswell, Bennett, and
  Lueck}{Zamani et~al\mbox{.}}{2020}]%
        {zamani2020generating}
\bibfield{author}{\bibinfo{person}{Hamed Zamani}, \bibinfo{person}{Susan
  Dumais}, \bibinfo{person}{Nick Craswell}, \bibinfo{person}{Paul Bennett},
  {and} \bibinfo{person}{Gord Lueck}.} \bibinfo{year}{2020}\natexlab{}.
\newblock \showarticletitle{Generating clarifying questions for information
  retrieval}. In \bibinfo{booktitle}{\emph{Proceedings of The Web Conference
  2020}}. \bibinfo{pages}{418--428}.
\newblock


\bibitem[\protect\citeauthoryear{Zamani, Trippas, Dalton, and Radlinski}{Zamani
  et~al\mbox{.}}{2022}]%
        {zamani2022conversational}
\bibfield{author}{\bibinfo{person}{Hamed Zamani}, \bibinfo{person}{Johanne~R
  Trippas}, \bibinfo{person}{Jeff Dalton}, {and} \bibinfo{person}{Filip
  Radlinski}.} \bibinfo{year}{2022}\natexlab{}.
\newblock \showarticletitle{Conversational information seeking}.
\newblock \bibinfo{journal}{\emph{arXiv preprint arXiv:2201.08808}}
  (\bibinfo{year}{2022}).
\newblock


\bibitem[\protect\citeauthoryear{Zhang, Jiang, Li, and Xue}{Zhang
  et~al\mbox{.}}{2020}]%
        {zhang-etal-2020-language-generation}
\bibfield{author}{\bibinfo{person}{Maosen Zhang}, \bibinfo{person}{Nan Jiang},
  \bibinfo{person}{Lei Li}, {and} \bibinfo{person}{Yexiang Xue}.}
  \bibinfo{year}{2020}\natexlab{}.
\newblock \showarticletitle{Language Generation via Combinatorial Constraint
  Satisfaction: A Tree Search Enhanced {M}onte-{C}arlo Approach}. In
  \bibinfo{booktitle}{\emph{Findings of the Association for Computational
  Linguistics: EMNLP 2020}}. \bibinfo{publisher}{Association for Computational
  Linguistics}, \bibinfo{address}{Online}, \bibinfo{pages}{1286--1298}.
\newblock
\urldef\tempurl%
\url{https://doi.org/10.18653/v1/2020.findings-emnlp.115}
\showDOI{\tempurl}


\bibitem[\protect\citeauthoryear{Zhang, Chen, Ai, Yang, and Croft}{Zhang
  et~al\mbox{.}}{2018}]%
        {zhang2018towards}
\bibfield{author}{\bibinfo{person}{Yongfeng Zhang}, \bibinfo{person}{Xu Chen},
  \bibinfo{person}{Qingyao Ai}, \bibinfo{person}{Liu Yang}, {and}
  \bibinfo{person}{W~Bruce Croft}.} \bibinfo{year}{2018}\natexlab{}.
\newblock \showarticletitle{Towards conversational search and recommendation:
  System ask, user respond}. In \bibinfo{booktitle}{\emph{Proceedings of the
  27th acm international conference on information and knowledge management}}.
  \bibinfo{pages}{177--186}.
\newblock


\bibitem[\protect\citeauthoryear{Zhao, Dou, Mao, and Wen}{Zhao
  et~al\mbox{.}}{2022}]%
        {zhao2022generating}
\bibfield{author}{\bibinfo{person}{Ziliang Zhao}, \bibinfo{person}{Zhicheng
  Dou}, \bibinfo{person}{Jiaxin Mao}, {and} \bibinfo{person}{Ji-Rong Wen}.}
  \bibinfo{year}{2022}\natexlab{}.
\newblock \showarticletitle{Generating Clarifying Questions with Web Search
  Results}. In \bibinfo{booktitle}{\emph{Proceedings of the 45th International
  ACM SIGIR Conference on Research and Development in Information Retrieval}}.
  \bibinfo{pages}{234--244}.
\newblock


\end{thebibliography}

\appendix

\section{Human Annotation Guideline}
\label{appendsec:guide}

In this task, imagine you are the user who unintentionally asks our search system an ambiguous search query (imagine they are using Google or talking to Siri) in a conversation. To better understand the intention of your query, our system asks a clarifying question to you. And your task is to judge if this clarification question is natural and useful. 

For example, the user asks "Tell me about defender". The query is ambiguous because the word "defender" can refer to a personality type coded as ISFJ, a TV series "The Defender", a vehicle named "Defender", or a video game named "Defender". In order to know whether the user is asking about the TV series, the search system asks a clarifying question "Are you interested in a television series?"

Another example can be the user asks "Tell me information about computer programming." Different from the last example, this query is NOT ambiguous because of the term "computer programming" is ambiguous, but because "computer programming" is a general concept, and there can be multiple search directions. For example, the user can be looking for computer programming jobs, computer programming languages, computer programming courses, or the history of computer programming. To confirm whether the user is looking for computer programming courses, the system asks a clarifying question "Are you looking for a course in computer programming?"

In general, ambiguous queries have many possible "facets". For example, "TV series" is one possible facet in the "defender" example, and "course" is one possible facet in the "computer programming" example. Our system generates these questions based on the ambiguous query and one possible facet. 

Your goal is to evaluate the clarifying question asked by our system, in terms of its Naturalness and Usefulness (Please read explanation below). Besides the query, facet, and generated question, you will also get a human-written question as your reference. You can assume the human-written question is always good in both naturalness and usefulness.

\textbf{Explanation of Naturalness}:

The Naturalness of a question is whether the question is fluent, grammatical, and easy to understand. Your goal is to give each question "Good", "Fair", or "Bad" in terms of its naturalness.
Good naturalness means the question is fluent and like our daily language. Fair naturalness means although not grammatically perfect or contains noise, the question can still be understood with efforts. Bad naturalness means the question is incomplete, hard to understand or the generated sentence is not a question.

Here are some examples with explanations:

Example 1

Query: "Tell me about defender"

Facet: "television series"

Reference: "are you interested in the television series defender"

Good naturalness questions:

"are you interested in a television series"  (Almost the same as reference)

"do you need to be in the team"  (fluent and easy to understand, although not meaningful)

Fair naturalness questions:

"do you want to know television series" (A little weird but understandable)

Bad naturalness questions:

"television series, etc." (Not a question, and the sentence is incomplete)

"would you like to know more about" (the question is incomplete)

Example 2

Query: "Tell me information about computer programming."

Facet: "courses"

Reference: "are you interested in coding courses online"

Good naturalness questions:

"are you looking for a course in computer programming"  (Almost the same as reference)

"do you need to have courses in computer science" (Almost the same as reference)

"would you like to tell me about it"  (fluent and easy to understand, although not meaningful)

Fair naturalness questions:

"do you want to coursework in computer programming" (sound strange/ungrammatical, but understandable on a second thought)

"do you want to know what is going on with your courses" (fluent but unlike daily language)

Bad naturalness questions:

"do you need to know" (Not a complete sentence)

\begin{figure}[t]
    \centering
    \includegraphics[width=\linewidth]{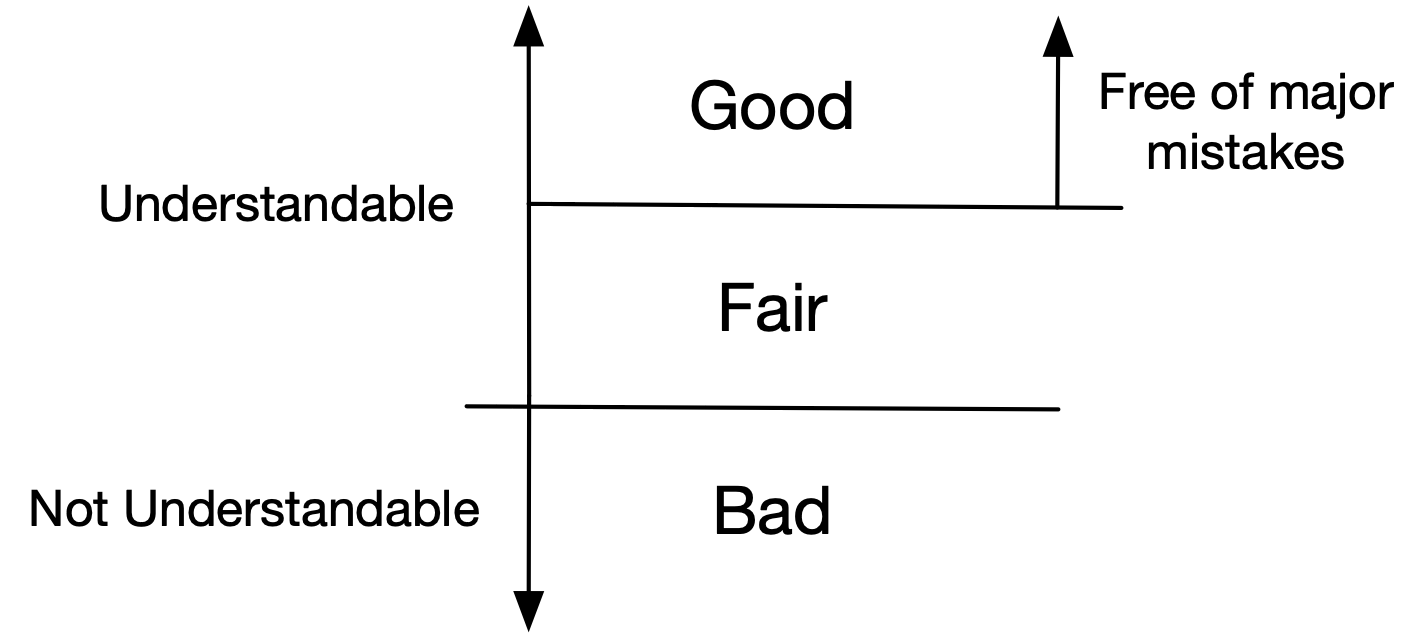}
    \caption{Decision Boundaries for Naturalness}
\end{figure}

\textbf{Explanation of Usefulness}:

 A useful question should:
 \begin{enumerate}
     \item Clarify the query if answered by the user.
     \item Not be a duplicative question of the query or ask for prequel information.
     \item Not miss the intent of the query.
     \item Not be too general or over-specific.
 \end{enumerate}
Your goal is to give each question "Good", "Fair", or "Bad" in terms of its usefulness.
A Good usefulness questions is a perfect reflection of the facet and make the query easier to answer. A Fair usefulness questions is weakly relevant to the query and facet, but not completely irrelevant. A Bad usefulness question can be completely irrelevant, duplicative, miss-intent, prequel, too general or over-specific.

A question can be natural but not useful. Please see examples below.

Example 1

Query: "Tell me about defender"

Facet: "television series"

Reference: "are you interested in the television series defender"

Good usefulness questions:

"are you interested in a television series"  (Almost the same as reference)

"do you want to know television series" (not perfectly natural but useful)

Fair usefulness questions:

"would you like to see a television series based on your work" (although "based on your work" is weird, user could still answer the question as "yes I am referring to the TV series defender")

Bad usefulness questions:

"do you need to be in the team"  (not meaningful for the query)

"television series, etc." (Not a question, and the sentence is incomplete, user cannot answer it)

Example 2

Query: "Tell me information about computer programming."

Facet: "courses"

Reference: "are you interested in coding courses online"

Good usefulness questions:

"are you looking for a course in computer programming"  (Almost the same as reference)

"do you need to have courses in computer science" (not natural but useful)

"do you want to coursework in computer programming" (not natural but useful)

Fair usefulness questions:

"do you want to know what is going on with your courses" (weakly relevant to the facet)

Bad usefulness questions:

"do you need to know" (Not a complete sentence)

"would you like to tell me about it"  (completely irrelevant)

\begin{figure}[t]
    \centering
    \includegraphics[width=\linewidth]{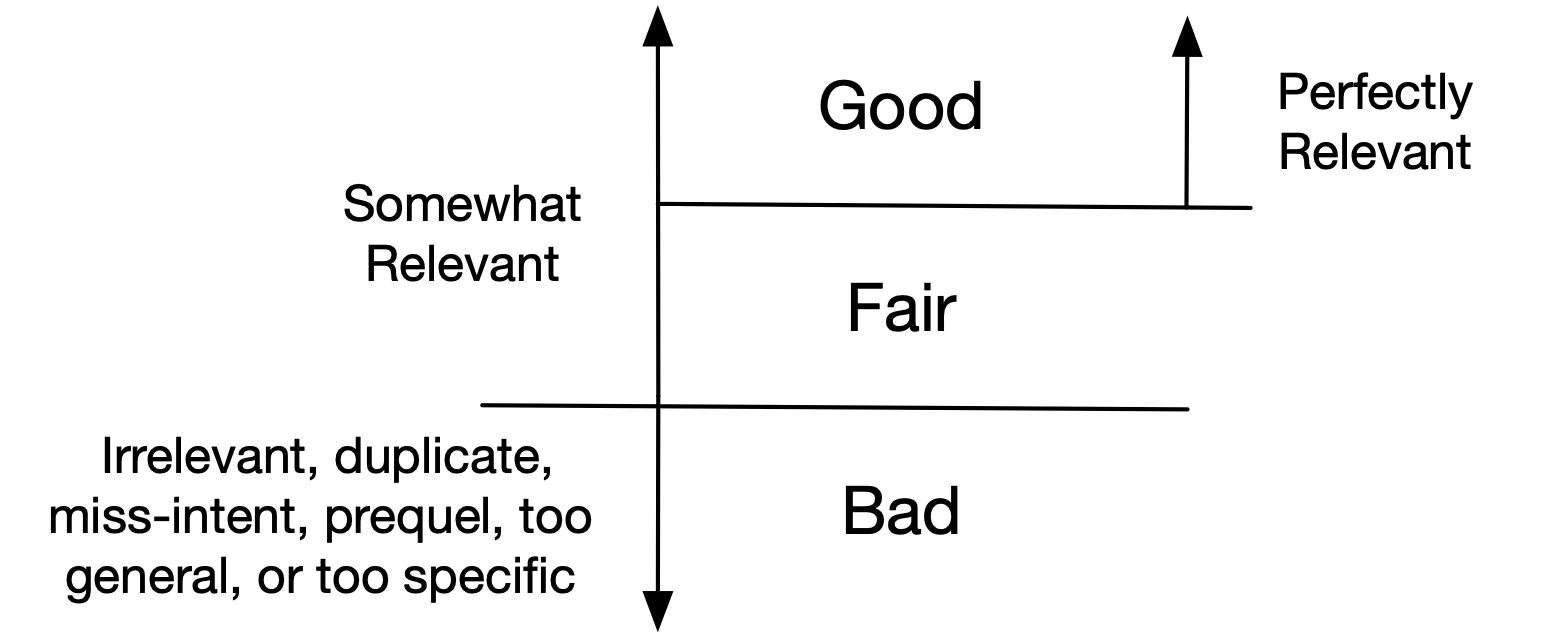}
    \caption{Decision Boundaries for Usefulness}
\end{figure}

\end{document}